\documentclass[epj,final,numbook]{svjour}

\usepackage{amsmath}
\usepackage{amssymb}
\usepackage{epsfig}
\usepackage{graphicx}
\usepackage{color}\usepackage[utf8]{inputenc}

\usepackage{amsfonts}
\usepackage{bbold}
\usepackage{mathptmx} 

\usepackage{bm}
\usepackage{xcolor}
\usepackage{fix-cm}
\usepackage[T1]{fontenc}

\usepackage{epstopdf}

\newcommand{\Ham}{\mathcal{H}}
\usepackage{parskip}

\usepackage[colorlinks,allcolors=blue]{hyperref}
\usepackage{cite}

\begin{document}

\title{Skyrme-quark-meson-coupling energy density functional predictions to nuclear ground state properties}

\author{E. McRae\inst{1}\thanks{\emph{ellen.mcrae@anu.edu.au}} \and C. Simenel\inst{1}\thanks{\emph{cedric.simenel@anu.edu.au}} }
\institute{Department of Fundamental and Theoretical Physics \& Department of Nuclear Physics and Accelerator Applications, Research School of Physics, The Australian National University, Canberra ACT  2601, Australia 
}
\date{\today}
\abstract{
We present a systematic study of nuclear ground-state properties obtained with the Skyrme quark-meson coupling (SQMC) energy density functional and compare them with results from the SLy4d Skyrme parameterisation. The SQMC functional is constructed using the quark–meson coupling (QMC) model, which incorporates the internal quark structure of the nucleon and results in a significant reduction in the number of free parameters. We investigate binding energies, two-nucleon separation energies, charge radii, and quadrupole deformations across a broad range of nuclei. Particular attention is devoted to the isovector dependence of the spin–orbit interaction derived within the QMC model, and its impact on the binding energies of neutron-rich nuclei relevant to the r-process. We find that the SQMC functional provides a reasonable description of nuclear ground-state properties in comparison with SLy4d.}

\maketitle

\section{\label{sec:Intro}Introduction}

Despite significant progress in applications of chiral effective field theory \cite{Weinberg1991,Holt2013} to `ab initio' nuclear structure and reaction calculations across light to medium-mass nuclei \cite{machleidt2024,Hildenbrand26}, 
energy density functional (EDF) approaches remain a popular tool to describe nuclear systems, from ground-states to excitations \cite{schunck2019,colo2020}, fission and heavy-ion reaction mechanisms \cite{simenel2018,schunck2022,simenel2025} across the whole nuclear chart. 
Self-consistent mean-field methods based on nuclear EDF, such as the Skyrme functional \cite{Skyrme1956}, render the calculation of properties of large numbers of nuclei computationally tractable.

Determining the parameters of an EDF, together with their associated uncertainties, remains a major challenge. One avenue aims at reducing the number of free parameters, for example by constraining the Skyrme functional within low-energy chiral QCD~\cite{Kaiser2010a,Kaiser2010b,Holt2011,Holt2013} or by fitting to ab initio nuclear and neutron matter equations of state~\cite{Baldo2013}. Another avenue emphasizes improvements in the EDF calibration protocol, as illustrated by the Universal Nuclear Energy Density Functional (UNEDF) parameterisations~\cite{Kortelainen2010,Kortelainen2012,kortelainen2014}.
\par
Alternatively, one may adopt a framework that explicitly incorporates the quark substructure of nucleons  and thus goes beyond a purely hadronic mean-field approximation. This approach was employed in \cite{chamseddine2023} with chiral potentials as well as in  the quark–meson coupling (QMC) model~\cite{Guichon1988,Guichon1996}. Within the QMC model, an EDF  is derived by exploiting the underlying quark degrees of freedom of the nuclear system through a self-consistent mean-field treatment of in-medium modifications to the quark structure of bound nucleons. Although it may be argued on energy-scale grounds that an explicit quark-level description is unnecessary, the QMC model does not a priori impose in-medium changes of the quark wave functions; rather, such modifications arise dynamically through their interactions with the mean meson fields, and are found to be significant. In particular, while the scalar and vector mean fields largely cancel at the hadronic level, only the scalar field—whose magnitude remains substantial—polarizes the quarks, thereby generating a novel density dependence.
\par
Of particular relevance, the QMC model predicts both the central and spin–orbit components of the effective nucleon–nucleon interaction, including their isovector structure\footnote{Recent implementations of the QMC model also predict tensor and pairing interactions~\cite{Martinez2020}.}. In this framework, the dominant contributions to the spin–orbit interaction arise naturally from Thomas precession and spatial variations of the vector mean fields ($\omega$ and $\rho$) across the finite size of the bound nucleon, leading to the emergence of anomalous isoscalar and isovector magnetic moments. Overall, the resulting EDF involves a relatively small number of free parameters, which can be calibrated to properties of finite nuclei~\cite{Stone2016,Martinez2019}. 
A comprehensive review of the QMC model can be found in Ref.~\cite{Guichon2018}.
\par
At present, numerous computational codes based on the Hartree-Fock approximation and its extensions are available for calculating nuclear structure and reaction properties (see, e.g., \cite{Dobaczewski2021,Marevic2022,Abhishek2024,ryssens2015,Reinhard2021}). The majority of these implementations rely on Skyrme EDF. 
 Although the QMC functional has been implemented directly into some of these solvers, including to solve the time-dependent HF equation \cite{godbey2019}, it is desirable to exploit the well-established Skyrme EDF framework.
To this purpose, the QMC model can be employed to constrain the parameters of a Skyrme functional, leading to a Skyrme-QMC functional (SQMC) as the one proposed by Wang et al~\cite{Wang2011}. Whereas conventional Skyrme EDFs typically involve ten or more adjustable parameters, the SQMC functional depends only on the four free parameters of the QMC model (excluding the pairing channel), {supplemented by a number of conventional modelling choices required for the mapping onto a Skyrme functional. This substantial reduction in the number of physics-driven parameters allows the SQMC functional to be constructed without fitting to experimental binding energies or charge radii of finite nuclei, in contrast to standard Skyrme EDFs, whose fitting protocols generally include such finite-nucleus data~\cite{Chabanat1998}.}

 It should be noted that, despite starting with quark degrees of freedom, QMC is a much simpler approach than QCD. In particular, QMC is not an effective field theory of QCD but rather a phenomenological approach aiming at describing the in-medium nuclear interaction. In addition, SQMC adds another layer of approximation by replacing the volume term density dependence with a Skyrme-like $\rho^\alpha$ ansatz.

The aim of this work is to present a systematic comparison of ground-state properties calculated using the SQMC functional and the SLy4d Skyrme parameterisation~\cite{SLy4d}. 
{ Our primary objective is to systematically assess the predictive performance of the SQMC functional for nuclear ground-state properties and to compare it with that of a conventional Skyrme parameterisation.}
In addition, we examine the impact of the isovector dependence of the spin–orbit interaction, as derived within the QMC model, on exotic nuclei and nuclei relevant to the r-process.
\par
This paper is organized as follows. In Sec.~\ref{sec:Skyrme}, we briefly recall the Skyrme functional, while the QMC model is outlined in Sec.~\ref{sec:QMCmodel}. The determination of the SQMC parameters is presented in Sec.~\ref{sec:SQMC}. Ground-state properties—including binding energies, two-nucleon separation energies, charge radii, and quadrupole deformations—are analyzed in Sec.~\ref{sec:Systematic}. The isovector dependence of the spin–orbit functional predicted by the QMC model, and its impact on the binding energies of exotic nuclei, including those relevant to the r-process, are discussed in Sec.~\ref{sec:IsovectorSO}. Finally, conclusions are drawn in Sec.~\ref{sec:Conclusion}.



\section{Skyrme EDF}\label{sec:Skyrme}
The energy density functional $E=\int \! d\bm{r}\,\Ham(\bm{r})$ is a fundamental ingredient to density functional theory  calculations in nuclear physics. 
For simplicity, pairing correlations are neglected in this section\footnote{The treatment of pairing is discussed in section \ref{sec:pairing}.} and time-reversal symmetry is assumed, i.e., restricting to ground states of even-even nuclei.
Under these assumptions,  the EDF is expressed as a functional of 
the local nucleon density 
\begin{equation}
\rho(\bm{r})= \sum_{isq} \varphi^{sq*}_{i}\varphi^{sq}_{i},
\end{equation} 
the kinetic energy density 
\begin{equation}
\tau(\bm{r})= \sum_{isq} \nabla\varphi^{sq*}_{i}\nabla\varphi^{sq}_{i},
\end{equation} 
and the gradient of the spin-orbit density 
\begin{equation}
\nabla\cdot\bm{J}(\bm{r})= -i\sum_{iss'q} \nabla\varphi^{sq*}_{i}\times\nabla\varphi^{s'q}_{i}\cdot\left\langle s\right|\bm{\sigma}\left|s'\right\rangle,
\label{eq:spin_orbit_density}
\end{equation} 
where the sums run over occupied single-particle states $\varphi_i$, with spin $s$ and isospin $q$.
\par
The EDF can be separated based on the occurrence of the densities and currents such that
\begin{equation}
    \Ham=\rho M_{N}+\frac{\tau}{2M_{N}}+\Ham_{0+3}+\Ham_{\textrm{eff}}+\Ham_{\textrm{fin}}+\Ham_{\textrm{SO}}.
    \label{eq:EDF}
\end{equation}
$\Ham_0+\Ham_3$ is a volume term responsible for saturation, $\Ham_{\textrm{eff}}$ is related to the effective nucleon mass, finite size effects are accounted for by the surface term $\Ham_{\textrm{fin}}$, and $\Ham_{SO}$ is induced by the spin-orbit interaction.
\par
Under time-reversal symmetry, the general Skyrme EDF takes the form
\begin{align}
    \Ham_{0+3}^{\textrm{Skyrme}}={}&\frac{1}{12}\Biggl\{\left[6t_{0}\left(1+\frac{x_{0}}{2}\right) + t_{3}\left(1+\frac{x_{3}}{2}\right)\rho^{\alpha}\right] \rho^{2} \Biggr. \nonumber \\
    &\Biggl.-\left[6t_{0}\left(x_{0}+\frac{1}{2}\right)+t_{3}\left(x_{3}+\frac{1}{2}\right)\rho^{\alpha}\right] \sum_{q=n,p}\rho_{q}^{2}\Biggr\}, \label{eq:H0H3Skyrme} \\
    \Ham_{\textrm{eff}}^{\textrm{Skyrme}}={}&\frac{1}{4}\Biggl\{\left[t_{1}\left(1+\frac{x_{1}}{2}\right)+t_{2}\left(1+\frac{x_{2}}{2}\right)\right]\rho\tau \Biggr.\nonumber \\
    &\Biggl.- \left[t_{1}\left(x_{1}+\frac{1}{2}\right)-t_{2}\left(x_{2}+\frac{1}{2}\right)\right]\sum_{q=n,p}\rho_{q}\tau_{q}\Biggr\}, \label{eq:HeffSkyrme}\\
    \Ham_{\textrm{fin}}^{\textrm{Skyrme}}={}&-\frac{1}{16}\Biggl\{\left[3t_{1}\left(1+\frac{x_{1}}{2}\right)-t_{2}\left(1+\frac{x_{2}}{2}\right)\right]\rho\nabla^{2}\rho \Biggr. \nonumber \\
    &\Biggl.- \left[3t_{1}\left(x_{1}+\frac{1}{2}\right)+t_{2}\left(x_{2}+\frac{1}{2}\right)\right]\sum_{q=n,p}\rho_{q}\nabla^{2}\rho_{q}\Biggr\}\textrm{,}\label{eq:HfinSkyrme}\\
    \Ham_{\textrm{SO}}^{\textrm{Skyrme}}={}&-\frac{1}{2}\left(W_{0}\rho\nabla\cdot\bm{J} + W_{0}'\sum_{q=n,p}\rho_{q}\nabla\cdot\bm{J}_{q}\right) . \label{eq:HsoSkyrme}
\end{align}
\par
First proposed by Reinhard and Flocard in 1995 \cite{Reinhard1995}, $W_{0}'$ is a degree of freedom added to some modern Skyrme functionals to improve the isovector dependence to the spin-orbit term. We will see in section~\ref{sec:QMCmodel} that an isovector spin-orbit term emerges naturally within the QMC model due to relativistic and finite-size effects.  
\par
The  parameters of the Skyrme EDF\footnote{ $\alpha$ is sometimes treated as a free parameter \cite{Kortelainen2012}.} ($x_{0-3}$, $t_{0-3}$, $W_{0}$, and $W_{0}'$) are typically determined by a fit to infinite nuclear matter properties and to the experimental masses and radii of a selection of atomic nuclei~\cite{Bender2003}. 

\section{The quark-meson coupling model}\label{sec:QMCmodel}
The quark-meson coupling model is a relativistic mean-field model for the interaction between nucleons.
Self-consistent treatment of quark polarisation within a nucleus yields an EDF, which can be used to constrain the parameters of the Skyrme functional.
\subsection{QMC formalism and assumptions}\label{sec:QMCtheory}
Proposed by Guichon in Ref.~\cite{Guichon1988}, the quark-meson coupling model is a relativistic mean-field approach to the nucleon-nucleon interaction.
A key distinction between the QMC model and traditional relativistic mean-field  \cite{Reinhard1989} approaches is that the QMC model derives the nucleon-nucleon interaction by coupling the meson fields to the quarks within the nucleons rather than to the nucleons themselves.
\par
In QMC, each nucleon is modelled as a spherical cavity (a `MIT bag') containing three confined quarks.
The assumption of non-overlapping bags will break down at some high density ($\approx 3\rho_{0}$ with $\rho_0\approx0.16$~fm$^{-3}$ the saturation density). 
However, it  is quite safe for the low to moderate densities of nuclear systems barely exceeding $\rho_0$~\cite{Guichon2006}.
\par
The quarks are coupled to  $\sigma$, $\omega$ and $\rho$ meson fields. Apart from the pion, these are the lightest mesons and as such they are the most important at the ranges relevant for low-energy nuclear physics. 
Recent versions of QMC also take pion exchange into account through a local-density approximation~\cite{Guichon2018}, resulting into a finite-range Fock term contribution to the mean-field. 
However, to preserve the simplicity of a local exchange term induced by Skyrme EDF, pion exchange is not included in this work. 
\par
The properties of the mesons are summarised in Table~\ref{tab:Mesons}. It is important to note that the $\sigma$ meson is not a real particle but rather an effective way to simulate the correlated exchange of two pions. As such, its mass is not well fixed by experiment and it remains as a free parameter of the theory \cite{Guichon2006}.
The QMC model, like traditional relativistic mean-field  approaches, treats mesons as elementary fields. Their composition, whether that is a quark-antiquark pair or more complicated multiple-meson states, is usually neglected. 
\begin{table}
    \caption{Summary of properties of the mesons included in this work.}
    \centering
    \begin{tabular}{c|cccl}
    \hline \\[-1.0em]
        Meson&$J^{P}$&Isospin&Mass (MeV)&Notes\\ 
        \hline
        $\sigma$&$0^{+}$&0&$600 - 750$&Medium-range attraction\\
        $\omega$&$1^{-}$&0&783&Short-range repulsion\\
        $\rho$&$1^{-}$&1&775&Isospin dependence\\ 
        \hline
    \end{tabular}
    \label{tab:Mesons}
\end{table}
\par
A key outcome of the QMC model is the emergence of an effective nucleon mass
\begin{equation}
M^{*}_{N}(\sigma(\bm{R}))=M_N-g_{\sigma}\sigma+\frac{d}{2}(g_{\sigma}\sigma)^2 
\label{eq:effM}
\end{equation}
where the bare nucleon mass, $M_{N}$, is modified by the coupling to the $\sigma$ meson field. The scalar polarisability, $d$, describes the response of the nucleon to an applied scalar field, and is well approximated by 
\begin{equation}
    d =0.0044+0.211R_B-0.0357R_B^2 \textrm{ (fm)},
\end{equation}
where $R_B$ is the bag radius~\cite{Guichon2006}.
This effective mass is the major point of difference between QMC and quantum hadrodynamics. Taking the ``heavy quark'' limit, $m_{q}\sim 300$~MeV, Eq.~\ref{eq:effM} reduces to the effective mass of quantum hadrodynamics, $M^{*}_{N}\approx M_N-g_{\sigma}\sigma$. However for the more realistic case of light quarks ($m_{q}\sim 5$~MeV), the internal structure of the nucleon has a significant effect~\cite{Saito1994}. 
\par
Several extensions to the QMC model have been developed which are beyond the scope of this work. In addition to the  pion exchange discussed above~\cite{Krein1999}, chirality has been treated  via a cloudy bag model~\cite{Nagai2008}, the quark nature of mesons has been included~\cite{Saito1997}, and tensor and  pairing interactions have been derived from the QMC model~\cite{Martinez2020}.
Over the course of the development of the original model and its variations, the QMC model has been  applied to the study of finite nuclei, hypernuclei, the EMC effect, and neutron stars~\cite{Guichon2018}.

\subsection{QMC EDF}\label{sec:QMCEDF}
The QMC model can be used to derive an energy density functional of a similar form to the Skyrme functional (Eq.~\ref{eq:EDF}) for use in low energy nuclear physics~\cite{Guichon1996,Guichon2006}.
\par
The contribution to the Hamiltonian acting on the quark fields is solved as a modified version of the free Dirac equation, assuming 3 quarks in their lowest energy state.
From this, the variation of the meson field within the nucleon bag is treated as a perturbation, which yields a spin-orbit interaction between the magnetic moment of the nucleon and the `magnetic' component
of the $\omega$ and $\rho$ meson fields.
The Lorentz transformation to the rest frame of the nucleus results in an additional spin-orbit contribution due to Thomas precession.
\par
Solving the equations of motion for the meson fields selfconsistently gives expressions for their energy in terms of the local nucleon densities.
This is then expressed as an EDF  with terms separated under the convention of Eq.~\ref{eq:EDF}:

\begin{eqnarray}
\Ham_{0+3}&=&\!\!\left(\!\frac{-3G_{\rho}}{32}+\frac{G_{\sigma}}{8\left(1+d\rho G_{\sigma}\right)^{3}}-\frac{G_{\sigma}}{2\left(1+d\rho G_{\sigma}\right)}+\frac{3G_{\omega}}{8}\!\right)\!\rho^{2} \nonumber\\
&&+\left(\frac{5G_{\rho}}{32}+\frac{G_{\sigma}}{8\left(1+d\rho G_{\sigma}\right)^{3}}-\frac{G_{\omega}}{8}\right)\left(\rho_{n}-\rho_{p}\right)^{2}, \label{eq:H0H3QMC} \\
\Ham_{\textrm{eff}}&=&\left(\frac{G_{\rho}}{4m_{\rho}^{2}}+\frac{G_{\sigma}}{2M_{N}^{2}}\right)\rho\tau\nonumber\\
&&+\left(\frac{-G_{\rho}}{8m_{\rho}^{2}}-\frac{G_{\sigma}}{2m_{\sigma}^{2}}+\frac{G_{\omega}}{2m_{\omega}^{2}}-\frac{G_{\sigma}}{4M_{N}^{2}}\right)\sum_{q=n,p}\rho_{q}\tau_{q}, \label{eq:HeffQMC}\\
\Ham_{\textrm{fin}}&=&\left(\frac{-3G_{\rho}}{16m_{\rho}^{2}}-\frac{G_{\sigma}}{2m_{\sigma}^{2}}+\frac{G_{\omega}}{2m_{\omega}^{2}}-\frac{G_{\sigma}}{4M_{N}^{2}}\right)\rho\nabla^{2}\rho \nonumber\\
&&+\left(\frac{9G_{\rho}}{32m_{\rho}^{2}}+\frac{G_{\sigma}}{8m_{\sigma}^{2}}-\frac{G_{\omega}}{8m_{\omega}^{2}}+\frac{G_{\sigma}}{8M_{N}^{2}}\right)\sum_{q=n,p}\rho_{q}\nabla^{2}\rho_{q}\label{eq:HfinQMC}\\
\Ham_{\textrm{SO}}&=&-\frac{1}{4M_{N}^{2}}\Biggl[\left(G_{\sigma}+G_{\omega}\left(2\mu_{s}-1\right)\right)\rho\nabla\cdot\bm{J}\Biggr. \nonumber\\
&&\!\!\!\!\!\!\!\!\!\!\!\!\!\!\!\!\!\!\!\!\!\!\!\!\!\!\!\!\!\!\!\!\!\Biggl.+\left(\frac{G_{\sigma}}{2}+\frac{G_{\omega}}{2}\left(2\mu_{s}-1\right)+\frac{3G_{\rho}}{8}\left(2\mu_{v}-1\right)\right)\sum_{q=n,p}\rho_{q}\nabla\cdot\bm{J}_{q}\Biggr],\label{eq:HsoQMC}
\end{eqnarray}
where
$
    G_{\sigma} = \frac{\left(3 g_{\sigma}^{q} \right)^2}{m_{\sigma}^2} \textrm{ , }
    G_{\omega} = \frac{\left(3 g_{\omega}^{q} \right)^2}{m_{\omega}^2} \textrm{ , and }
    G_{\rho} = \frac{\left(3 g_{\rho}^{q} \right)^2}{m_{\rho}^2}\textrm{.}    
$
\par
One assumes experimental values for the free nucleon mass ($M_{N}$), the $\omega$ and $\rho$ meson masses ($m_{\omega}$ and $m_{\rho}$), and the isoscalar and isovector nucleon magnetic moments ($\mu_{s}$ and $\mu_{v}$).
This leaves us with four free parameters for the model: the $\sigma$ meson mass ($m_{\sigma}$), and a coupling constant for each of the three meson fields ($G_{\sigma}$, $G_{\omega}$ and $G_{\rho}$). 
\par
Thus, the QMC model has yielded a non-relativistic, zero-range effective interaction producing an EDF of a similar form to Skyrme, except for the density dependence in $\Ham_{0+3}$.

\section{Skyrme-QMC}
\label{sec:SQMC}
The Skyrme QMC (SQMC) is  a Skyrme parameterisation inspired by the QMC model. 
SQMC uses the usual $\rho^\alpha$ density dependence of Skyrme EDF rather than the more complicated QMC one, allowing for straightforward implementation in  (time-dependent) Hartree-Fock (HF) solvers using Skyrme EDF to evaluate the nuclear mean-field. 
As in QMC, the  free parameters of SQMC are the coupling constant for each of the three meson fields ($G_{\sigma}$, $G_{\omega}$ and $G_{\rho}$) and the $\sigma$ meson mass ($m_{\sigma}$). 
{ In addition, the mapping from QMC to Skyrme EDF requires fixing the coefficient $\alpha$ of the density dependence. Values of $\alpha$ ranging between $\frac{1}{6}$ and $\frac{1}{3}$  are commonly used \cite{Bender2003}, while the UNEDF functionals leave $\alpha$ as a free parameter \cite{Kortelainen2010}. By expanding the $\mathcal{H}_{0+3}$ term from QMC [see (Eq.~\ref{eq:H0H3QMC})] around the saturation density and by studying the variation of the symmetry energy with changing the parameter $\alpha$, Wang {\it et al.} \cite{Wang2011} showed that a value of about $\frac{1}{6}$ was adequate. Following their work, we then choose $\alpha=\frac{1}{6}$. 
}
However, Wang et al \cite{Wang2011}  fixed $W_{0}=W_{0}'$, while the SQMC parameterisation described below does not impose this restriction on the spin-orbit term. 
A density dependent zero-range pairing interaction is added in the usual way (see Sec. \ref{sec:pairing}).
\subsection{QMC coupling constants}\label{sec:QMCconstants}
\par
As in Ref.~\cite{Guichon2006}, the meson coupling constants are fixed using infinite nuclear matter pseudodata\footnote{ 
Although these values are representative empirical nuclear matter properties commonly adopted in the literature, they vary slightly from modern estimates. See, e.g., Ref.~\cite{drischler2024} that recommends $\rho_0\approx0.157\pm0.011$~fm$^{-3}$, $e_\infty\approx-15.97\pm0.40$~MeV and $a_S\approx32.0\pm1.1$~MeV. Nevertheless, the values in Eqs.~(\ref{eq:pseudodatarho}), (\ref{eq:pseudodataE}) and (\ref{eq:pseudodataA}) are used to facilitate the comparison with the QMC calculations of Ref.~\cite{Guichon2006}.}
 evaluated at saturation~\cite{SkM*},
\begin{align} 
    \textrm{nucleon density, }\rho_{0}&=0.16 \textrm{ fm}^{-3},\label{eq:pseudodatarho}\\
    \textrm{binding energy per nucleon, }e_{\infty}&=-15.85 \textrm{ MeV, and}\label{eq:pseudodataE}\\
    \textrm{symmetry energy, }a_{s}&=30 \textrm{ MeV.} \label{eq:pseudodataA}
\end{align}
\par
The $\sigma$ meson is not a physical particle, but is included to simulate correlated two-pion exchange, thus its mass is not fixed by experiment.
As discussed in Ref.~\cite{Guichon2006}, a choice of $m_{\sigma}$ in the range of 600-750~MeV is considered reasonable.
In particular, choosing $m_{\sigma}=700$ MeV and a nucleon bag radius $R_{B}=0.8$ fm gives values for the three meson coupling constants \cite{Guichon2006}
\begin{align}
    G_{\sigma}&=12.254 \textrm{ fm}^2,\\
    G_{\omega}&=8.899 \textrm{ fm}^2,\\
    G_{\rho}&=7.724 \textrm{ fm}^2.
    \label{eq:CouplingConstants}
\end{align}
\par
\begin{figure}
    \begin{center}
    \includegraphics[width=0.9\columnwidth]{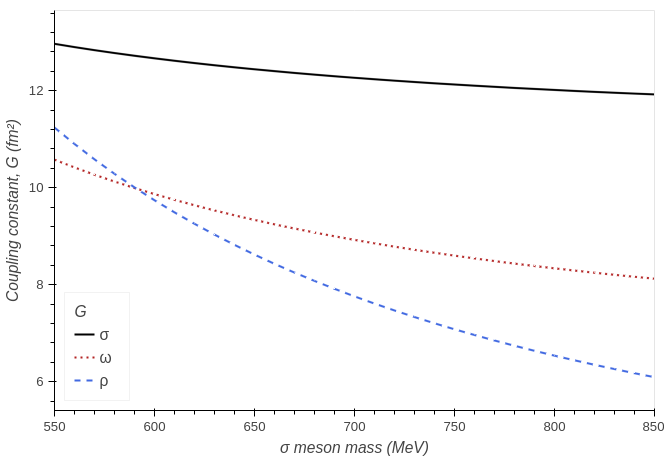}
    \end{center}
    \caption{\label{fig:coupling_constants}Meson coupling constants, $G_{\sigma,\omega,\rho}$, as a function of the $\sigma$ meson mass, $m_{\sigma}$.}
\end{figure}
The values for the meson coupling constants are relatively stable to the input data. As shown in Figure~\ref{fig:coupling_constants}, changing $m_{\sigma}$ by 100~MeV alters the $\sigma$ and $\omega$ coupling constants by an average of 2\% and 5\%, respectively.
$G_{\rho}$ varies more rapidly than $G_{\sigma,\omega}$. This is an indication that it will be important to more tightly constrain the value of $m_{\sigma}$ for the study of exotic systems, where isovector effects are more significant.

Ref.~\cite{Stone2016} uses the same version of the QMC model described in this work, but fixes the free parameters of the model through a fit to ground state properties of nuclei. The fit results in values of $G_{\sigma,\omega,\rho}$ and $m_\sigma$ that differ from those above by 1-7\%.

\subsection{Fitting protocol}\label{sec:SQMCfitting}
The construction of SQMC parameterisation primarily follows a process of adjusting the Skyrme parameters by equating terms of the Skyrme EDF (Eqs.~\ref{eq:H0H3Skyrme}~--~\ref{eq:HsoSkyrme}) to terms of the QMC EDF (Eqs.~\ref{eq:H0H3QMC}~--~\ref{eq:HsoQMC}). 
However, the volume terms (Eqs.~\ref{eq:H0H3Skyrme} and \ref{eq:H0H3QMC}) cannot be simply equated, as the QMC version has a more complex density dependence. With a choice of $\alpha=1/6$, a numerical fit over the region $\rho\in[0.12, 0.2\textrm{ fm}^{-3}]$ (see Appendix \ref{sec:fit_sensitivity}) is performed for symmetric matter to determine $t_{0,3}$, while $x_{0,3}$ are fit to infinite nuclear matter with $Z/A=82/208$ (corresponding to $^{208}$Pb) over the same density range. 
\begin{figure}
    \begin{center}
    \includegraphics[width=0.8\columnwidth]{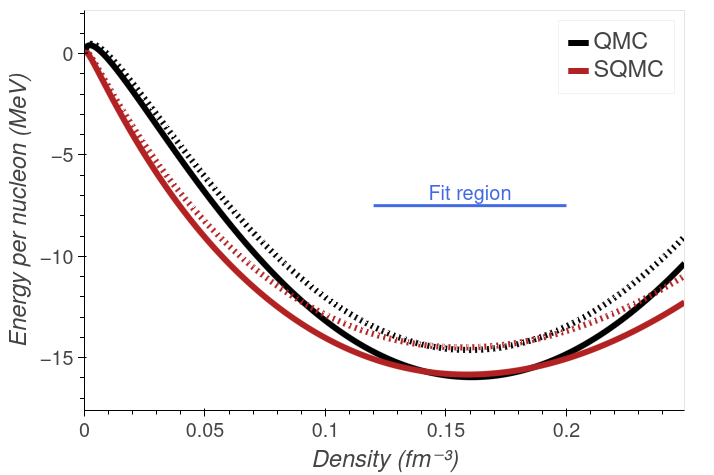}
    \end{center}
    \caption{\label{fig:SQMC_fit}Equations of state for the QMC and SQMC functionals. Solid lines: Symmetric nuclear matter. Dotted lines: Asymmetric nuclear matter corresponding to $Z/A=82/208$.}
\end{figure}

Figure~\ref{fig:SQMC_fit} compares the equations of state for the QMC and SQMC functionals showing the different density dependence. 
SQMC has a much lower incompressibility ($K_\infty^{SQMC}\simeq218.7$~MeV) compared to QMC ($K_\infty^{QMC}\simeq346$~MeV \cite{Guichon2006}). 
It should be noted that more recent versions of QMC including pions and $\sigma$ self-interaction lead to a much smaller incompressibility of the order of 230~MeV \cite{Martinez2019}. Further discussion of volume term fitting is included in appendix \ref{sec:fit_sensitivity}.

The  $\Ham_{\textrm{eff}}^{\textrm{QMC}}$, $\Ham_{\textrm{fin}}^{\textrm{QMC}}$ and $\Ham_{\textrm{SO}}^{\textrm{QMC}}$ terms of the QMC functional can be exactly reproduced by the Skyrme EDF (Eqs.~\ref{eq:H0H3Skyrme}~--~\ref{eq:HsoSkyrme}) for all densities and nucleon asymmetries by solving 
 a system of 4 equations for $t_{1,2}$ and $x_{1,2}$:
\begin{align}
    \frac{1}{4}\left[t_{1}\left(1+\frac{x_{1}}{2}\right)+t_{2}\left(1+\frac{x_{2}}{2}\right)\right] ={}&\left(\frac{G_{\rho}}{4m_{\rho}^{2}}+\frac{G_{\sigma}}{2M_{N}^{2}}\right) , \label{eq:sol1}\\
    -\frac{1}{4}\left[t_{1}\left(x_{1}+\frac{1}{2}\right)-t_{2}\left(x_{2}+\frac{1}{2}\right)\right] ={}&\left(\frac{-G_{\rho}}{8m_{\rho}^{2}}-\frac{G_{\sigma}}{2m_{\sigma}^{2}}\right.\nonumber\\
    &\left.+\frac{G_{\omega}}{2m_{\omega}^{2}}-\frac{G_{\sigma}}{4M_{N}^{2}}\right) , \label{eq:sol2}\\
    -\frac{1}{16}\left[3t_{1}\left(1+\frac{x_{1}}{2}\right)-t_{2}\left(1+\frac{x_{2}}{2}\right)\right] ={}&\left(\frac{-3G_{\rho}}{16m_{\rho}^{2}}-\frac{G_{\sigma}}{2m_{\sigma}^{2}}\right.\nonumber\\
    &\left.+\frac{G_{\omega}}{2m_{\omega}^{2}}-\frac{G_{\sigma}}{4M_{N}^{2}}\right) ,  \label{eq:sol3}\\
    \frac{1}{16}\left[3t_{1}\left(x_{1}+\frac{1}{2}\right)+t_{2}\left(x_{2}+\frac{1}{2}\right)\right]={}&\left(\frac{9G_{\rho}}{32m_{\rho}^{2}}+\frac{G_{\sigma}}{8m_{\sigma}^{2}}\right.\nonumber\\
    &\left.-\frac{G_{\omega}}{8m_{\omega}^{2}}+\frac{G_{\sigma}}{8M_{N}^{2}}\right). \label{eq:sol4}
\end{align}
The spin-orbit parameters are given by
\begin{align}
    W_{0}={}&\frac{1}{2M_{N}^{2}}\left(G_{\sigma}+G_{\omega}\left(2\mu_{s}-1\right)\right) , \label{eq:W0}\\
    W_{0}'={}&\frac{1}{4M_{N}^{2}}\left(G_{\sigma}+G_{\omega}\left(2\mu_{s}-1\right)+\frac{3G_{\rho}}{4}\left(2\mu_{v}-1\right)\right) . \label{eq:W0p}
\end{align}
The properties of this SQMC parameterisation are summarised in Tab.~\ref{tab:INMproperties} and its parameters are provided in Tab.~\ref{tab:Parameters}.
\begin{table}
    \caption{\label{tab:INMproperties}Properties of symmetric nuclear matter: saturation density $\rho_{0}$ (fm$^{-3}$), energy per nucleon at saturation $e_{\infty}$ (MeV), incompressibility $K_{\infty}$ (MeV), isoscalar effective nucleon mass $m^{*}_{s}$, enhancement factor of the Thomas-Reiche-Kuhn sum rule $\kappa$ (related to the isovector effective mass), and symmetry energy $a_{s}$ (MeV).}
    \centering
    \begin{tabular}[h]{llllll}
    \hline\\[-1.0em]
    $\rho_0$ & $e_{\infty}$ & $K_{\infty}$  & $m^{*}_{s}/M_{N}$ & $\kappa$& $a_{s}$ \\
    \hline
    0.1588               & -15.83             & 218.7              & 0.7727            & 0.5980  & 29.87 \\
     \hline
    \end{tabular}
\end{table}
\begin{table}
    \caption{\label{tab:Parameters}SQMC parameters.}
    \centering
    \begin{tabular}{l|l}
    Parameter& Value\\
    \hline
    $t_{0}$ (MeV fm$^{3}$)             & -2643.869 \\
    $t_{1}$ (MeV fm$^{5}$)             & 370.829  \\
    $t_{2}$ (MeV fm$^{5}$)             & -121.666  \\
    $t_{3}$ (MeV fm$^{3+3\alpha}$) & 15490.99 \\
    $x_{0}$                            & 0.59761   \\
    $x_{1}$                            & 0.266456  \\
    $x_{2}$                            & -0.227015 \\
    $x_{3}$                            & 0.693348  \\
    $\alpha$                       & $1/6$    \\
    $W_{0}$ (MeV fm$^{5}$)             & 82.8602   \\
    $W_{0}'$ (MeV fm$^{5}$)            & 147.4411  \\
    \end{tabular}
\end{table}

 \subsection{Doubly magic nuclei}

Ground-states of doubly-magic nuclei can be computed with the Skyrme and QMC EDF without pairing correlations at the mean-field level. Tables~\ref{tab:magicBE} and~\ref{tab:magicrc}  provide a summary of binding energies per nucleon and charge radii, respectively.  The QMC results were obtained in \cite{Guichon2006} with the same parameters as those used here for SQMC. The only difference comes from the density dependence in $\Ham_{0+3}$ [compare Eqs.~(\ref{eq:H0H3Skyrme}) and~(\ref{eq:H0H3QMC})].

As can be seen in Tab~\ref{tab:magicBE}, SQMC binding energies are very close to those obtained with SLy4d. Both interactions are in excellent agreement with experiment, except for $^{16}$O which is slightly underbound. However, smaller binding energies are found in the QMC calculations of \cite{Guichon2006}, which is likely due to the large incompressibility in the early version of QMC \cite{Martinez2019}. 

\begin{table}[h]
    \caption{\label{tab:magicBE} Comparison of binding energies per nucleon $e_\infty$ (MeV) for magic nuclei. QMC values are from \cite{Guichon2006}. SQMC and SLy4d values were obtained  with the \textsc{SkyAx} solver \cite{Reinhard2021}.}
    \centering
    \begin{tabular}{l|llll}
            & QMC \cite{Guichon2006}    & SQMC   & Sly4d  & Exp.  \\
    \hline
$^{16}$O    & 7.618   & 7.566  & 7.593  & 7.976 \\
$^{40}$Ca   & 8.213   & 8.531  & 8.557  & 8.551 \\
$^{48}$Ca   & 8.343   & 8.617  & 8.638  & 8.666 \\
$^{208}$Pb  & 7.515   & 7.88   & 7.865  & 7.867 \\
    \end{tabular}
\end{table}

{Table \ref{tab:magicrc} compares  experimental charge radii $r_c$ with theoretical predictions from QMC, SQMC and SLy4d. 
All three approaches reproduce the charge radii of doubly magic nuclei reasonably well. SQMC generally improves upon the original QMC values, while the SLy4d results remain of comparable quality.}

\begin{table}[h]
    \caption{\label{tab:magicrc} Same as Tab.~\ref{tab:magicBE} for charge radii $r_c$ (fm)}
    \centering
    \begin{tabular}{l|llll}
            & QMC \cite{Guichon2006}    & SQMC   & Sly4d  & Exp.  \\
    \hline
$^{16}$O    & 2.702   & 2.769  & 2.828  & 2.73  \\
$^{40}$Ca   & 3.415   & 3.482  & 3.544  & 3.485 \\
$^{48}$Ca   & 3.468   & 3.522  & 3.564  & 3.484 \\
$^{208}$Pb  & 5.42    & 5.506  & 5.556  & 5.5   \\
    \end{tabular}
\end{table}

\subsection{Pairing correlations}\label{sec:pairing}

\par
Although a pairing interaction can be derived within the QMC model~\cite{Martinez2020}, here we adopt the simple ``mixed'' pairing effective interaction
\begin{equation}
V_{\textrm{pair}}(\bm{r},\bm{r}') = V_{0}\left(1-\frac{\rho(\bm{r})}{2\rho_{0}}\right)\delta\left(\bm{r}-\bm{r}'\right),
\label{eq:pairing_int}
\end{equation}
where $V_{0}$ is the pairing strength parameter, $\rho(\bm{r})$ is the nucleon density, and $\rho_{0}$ is the saturation density.
\par
The pairing gap for state $k$ is given by
\begin{align}
    \Delta_k = \int d\bm{r} |\varphi_{k}|^2 \Delta (\bm{r})\label{eq:pairing_gap_k}
\end{align}
where $\Delta(\bm{r})$ is the pairing field.
An average pairing gap can be defined from this, weighted by either the occupation probability, $v_k^2$, or occupation amplitudes, $u_k v_k$, as~\cite{Bender2003}
\begin{align}
    \Delta_q &= \frac{\sum_k v_k^2\Delta_k}{\sum_k v_k^2} \textrm{ or,}\label{eq:pairing_gap_v2}\\
    \Delta_q &= \frac{\sum_k u_k v_k \Delta_k}{\sum_k u_k v_k} \textrm{,}\label{eq:pairing_gap_uv}
\end{align}
with the latter definition more sensitive to states near the Fermi surface.
A commonly used approach is to adjust the pairing strength to result in an average neutron pairing gap of $\Delta_n=1.245$~MeV for $^{120}$Sn~\cite{Dobaczewski2001}.
See Appendix \ref{sec:codes} for more details on the pairing strength parameter fitting. 

\subsection{SQMC input summary}

To make the notion of parameter freedom more transparent, we separate the different classes of inputs entering the SQMC construction. The inputs of the underlying QMC model are the $\sigma$-meson mass $m_\sigma$, the couplings $G_{\sigma,\omega,\rho}$, and the bag radius $R_B$; in this work we take $m_\sigma = 700$ MeV (within the typical 600--750 MeV range) and $R_B = 0.8$ fm, from which the couplings are fixed. These are determined using nuclear matter properties at saturation ($\rho_0 = 0.16\,\mathrm{fm}^{-3}$, $e_\infty = -15.85$ MeV, $a_s = 30$ MeV). 

The mapping onto a Skyrme-like form then requires standard technical choices, such as $\alpha = 1/6$ and a density interval $\rho \in [0.12, 0.2]\,\mathrm{fm}^{-3}$ for the fit of the symmetric and asymmetric (with $Z/A=82/208$ as in $^{208}$Pb) nuclear matter. In addition, pairing is treated separately through a density-dependent delta interaction, with its strength adjusted to reproduce the neutron gap $\Delta_n=1.245$~MeV  in $^{120}$Sn, as is customary in Skyrme calculations. 
This separation highlights that SQMC involves a small set of physics-driven inputs supplemented by conventional EDF modelling choices.

\section{Systematic study of ground state properties}
\label{sec:Systematic}
One notable advantage of developing a Skyrme-based functional like SQMC is that it can easily be included in existing HF codebases.
See Appendix \ref{sec:codes} for a discussion of the \textsc{SkyAx} \cite{Reinhard2021} and \textsc{hfbtho} \cite{Stoitsov2013} solvers used in this work.

A broad systematic study of ground state properties is  performed for all even-even nuclei with a known mass, and with proton and neutron numbers greater than 7.
These properties include binding energy, two-particle separation energies, charge radii, and quadrupole deformation.
Unless otherwise stated, all calculations reported in this section use \textsc{SkyAx}, as it offers a good balance of calculation speed and broad applicability across the nuclear chart. 
\par
The performance of the SQMC functional, developed in Chapter~\ref{sec:SQMC}, will be compared to a traditional Skyrme functional, SLy4d \cite{SLy4d}.
The latter, based on SLy4~\cite{Chabanat1998}, was created without centre-of-mass corrections\footnote{The fitting of SQMC parameters does not require HF calculations of finite nuclei. SQMC can thus be used without centre of mass corrections.} so that it could be consistently applied to both static and time-dependent calculations (`d' stands for dynamics).


\subsection{Binding energy}
The accurate reproduction of ground state binding energies of nuclei is a fundamental benchmark to assess functionals used in Hartree-Fock approaches.
The experimental data used for comparison in this section is the AME2020~\cite{Wang2021} evaluation.
Note that AME2020 includes a number of `estimated' mass values that are partly derived from trends in the masses of surrounding nuclei, which are included in this work.
\par
\begin{figure}[!htbp]
    \centering
    \includegraphics[width=0.9\columnwidth]{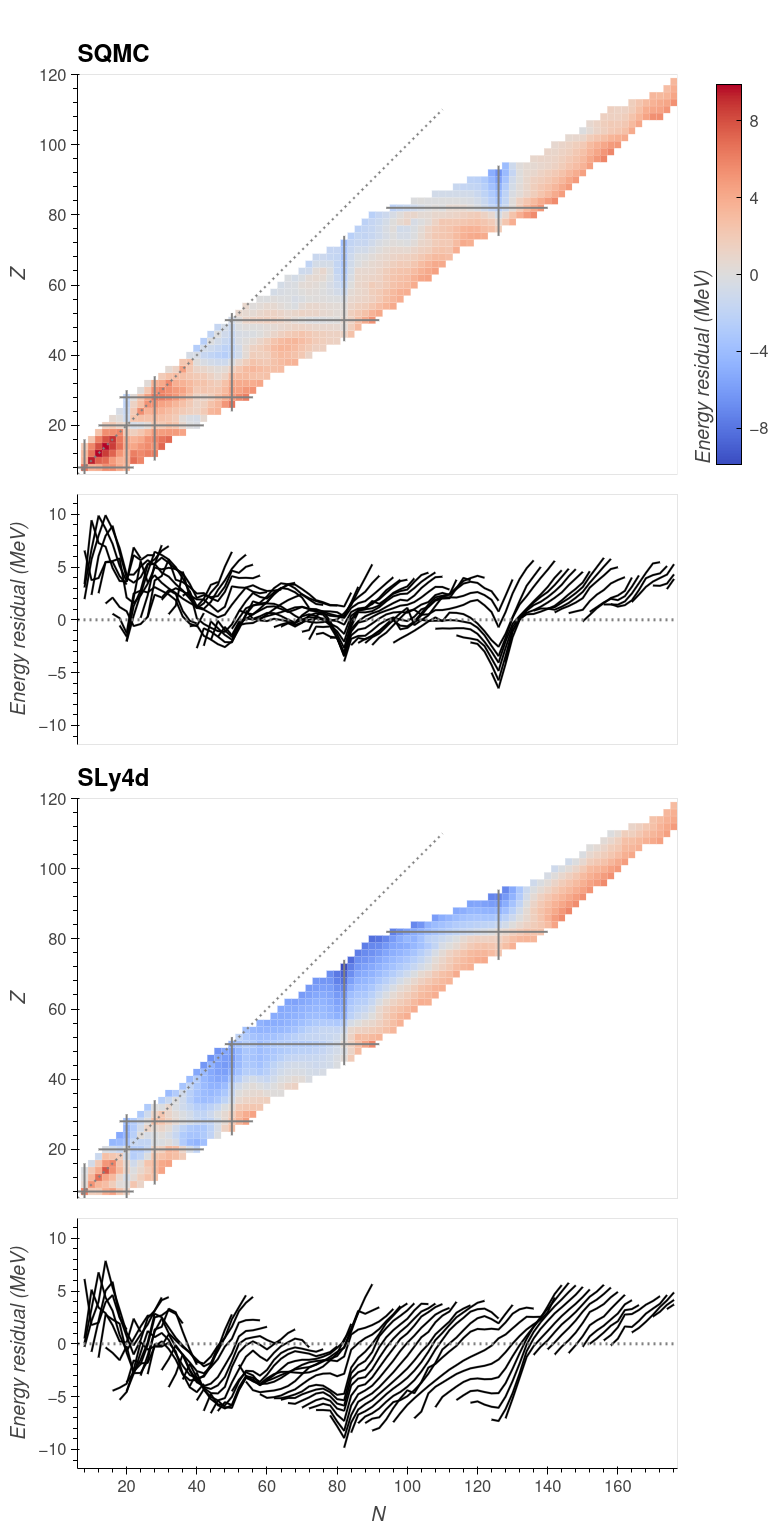}
    \caption{Binding energy residuals, $E_{\textrm{cal}} - E_{\textrm{exp}}$, obtained with \textsc{SkyAx} \cite{Reinhard2021}. Upper: SQMC. Lower: SLy4d. In the 2\textsuperscript{nd} and 4\textsuperscript{th} panels, each line corresponds to an isotopic chain.}
    \label{fig:BE_residuals_SQMC_SLy4d}
\end{figure}
Figure~\ref{fig:BE_residuals_SQMC_SLy4d} shows binding energy residuals relative to the experimental values across the nuclear chart for the SQMC and SLy4d functionals.
In the 1\textsuperscript{st} and 3\textsuperscript{rd} panels, blue indicates nuclei where the calculation is more strongly bound than experiment, and red indicates the opposite. 
The 2\textsuperscript{nd} and 4\textsuperscript{th} panels collapse this same residual data onto a single axis (neutron number) with lines corresponding to each isotopic chain.
\par
 It is interesting that SLy4d binding energy residuals do not exhibit the overall increase with mass as observed with SLy4 \cite{bertsch2005,dobaczewski2004}. This could be due to the different treatment of centre-of-mass corrections between SLy4 and SLy4d which could in turn impact nuclear matter properties \cite{dacosta2024}.
For mid-to-high mass region of $Z$ above 50, SLy4d shows a distinct tendency to overestimate masses for the lighter nuclei of an isotopic chain, and underestimate for heavier nuclei (i.e. blue to red across an isotopic chain). 
The SQMC predictions for this region display a  weaker dependence on $N$, becoming more pronounced for nuclei above $Z=82$.
\par
Overall, SLy4d tends to overestimate the strength of binding, while SQMC tends to underestimate, substantially so in the case of light nuclei.
Looking at only nuclei with $N<28$ and $Z<28$, the average binding energy residual for SQMC is 4.17 MeV, while SLy4d is much more centred with a mean of 0.78 MeV. Both functionals give a similarly broad spread of values for this region. The standard deviation of binding energy residuals for $N,Z<28$ is 2.78~MeV and 2.85~MeV for SQMC and SLy4d, respectively.
\par
At the other end of the nuclear chart, if we consider the area around the $N=126$ shell closure we see a clear band of excess binding in the predictions of SQMC. The 2\textsuperscript{nd} panel of Fig.~\ref{fig:BE_residuals_SQMC_SLy4d} shows that SQMC has a distinct turning point at $N=126$ for each isotopic chain. This may be an indication that spherical shell effects at $N=126$ are too strong in SQMC. The residuals for SLy4d on the other hand, have a much smoother transition across this region.
\par
Considering the full nuclear chart, it appears that SQMC results in a narrower spread of residuals than SLy4d, but that the predictions are more biased towards underbinding. This is borne out by the mean and standard deviation of binding energy residuals. The residuals for SQMC are centred around 1.58~MeV with a spread of 2.30~MeV, while SLy4d has a mean of -0.89~MeV and standard deviation of 3.28~MeV. For the full nuclear chart, the binding energy predictions of SLy4d are more centred on the experimental values but with a broader spread than SQMC.
\par
In addition to considering the mean of the residuals, the data of Figure~\ref{fig:BE_residuals_SQMC_SLy4d} can be summarised as a root mean square (RMS) of the residuals defined by
\begin{align}
    \textrm{RMS residual} = \sqrt{\frac{1}{N}\sum_{i=1}^{N}\left(E_{\textrm{cal}} - E_{\textrm{exp}}\right)^{2}}.
\end{align}
The RMS residual of binding energy for SQMC is 2.79~MeV, which is slightly better than that of SLy4d at 3.40~MeV.
As shown in Table~\ref{tab:RMS_energy_residuals}, this is consistent with systematic studies with other functionals. Note that these residuals are computed with different pairing methods and solvers leading to some variations as can be seen from the two UNEDF1 studies. 
Note  the small residual obtained with QMC$\pi$-III-T, the latest version of QMC  including pions, $\sigma$ self interaction, tensor and pairing \cite{Martinez2020}. However, the fitting protocol of QMC$\pi$-III-T includes data from finite nuclei.
\begin{table}
    \caption{\label{tab:RMS_energy_residuals}RMS residual of binding energy for selected functionals. The SQMC and SLy4d values are calculated using \textsc{SkyAx} \cite{Reinhard2021} with BCS pairing. The UNEDF1 value with HFB pairing method is from Ref.~\cite{Kortelainen2012} and was obtained with \textsc{hfbtho} \cite{Stoitsov2013}. The UNEDF1 and SV-min~\cite{SVmin} values with BCS are from Ref.~\cite{Martinez2019} and were obtained with \textsc{SkyAx}. The  DD-ME$\delta$ value was obtained with the relativistic Hartree-Bogoliubov (RHB) approach in Ref.~\cite{DDMEdelta}.  QMC$\pi$-III-T is the most recent QMC functional. Its value is from \cite{Martinez2020} and was obtained with a modified version of \textsc{SkyAx}.}
    \centering
    \begin{tabular}{lll}
        Functional&RMS residual (MeV)& Pairing method\\
        \hline
        SQMC&2.79 (this work)&BCS \\
        SLy4d&3.40 (this work)&BCS \\
        UNEDF1&1.91 \cite{Kortelainen2012} & HFB\\
        UNEDF1&2.12 \cite{Martinez2019} & BCS\\
        SV-min&3.10 \cite{Martinez2019} & BCS\\
        DD-ME$\delta$& 2.33 \cite{DDMEdelta}&RHB\\
        QMC$\pi$-III-T& 1.74 \cite{Martinez2020} & BCS\\
    \end{tabular}
\end{table}
\par
Together the RMS residuals and Figure~\ref{fig:BE_residuals_SQMC_SLy4d} show that, when compared to SLy4d, the SQMC functional delivers comparable performance in reproducing experimental binding energies over the whole nuclear chart.

\subsection{Two-particle separation energies}
\par
\begin{figure}[!htb]
    \centering
    \includegraphics[width=0.9\columnwidth]{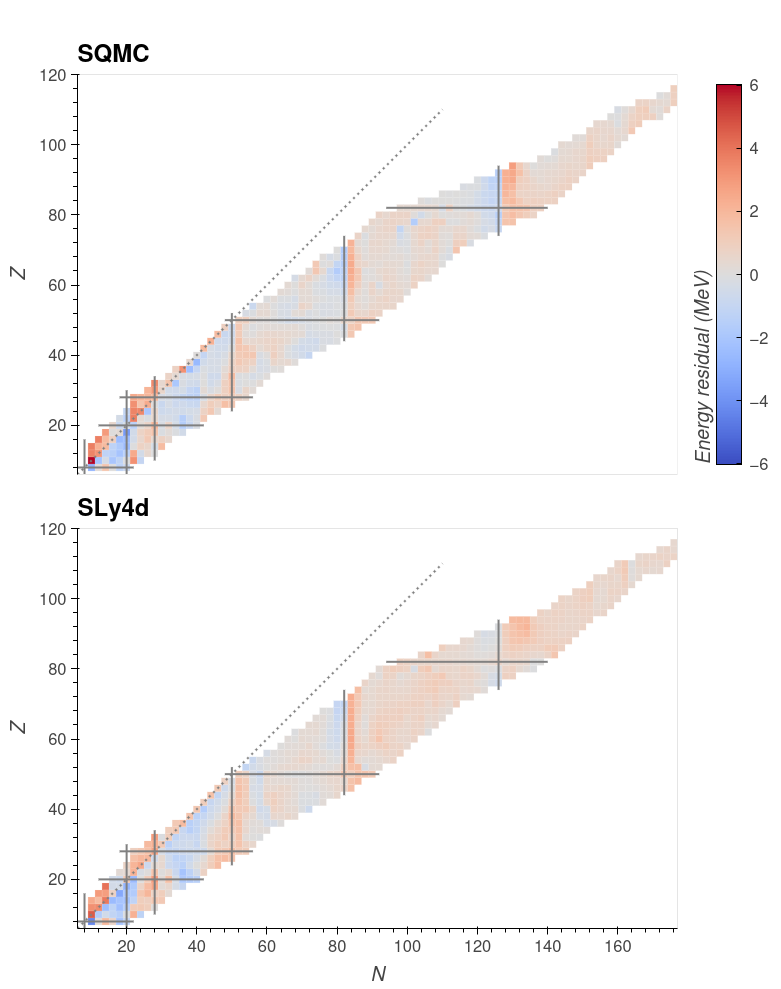}
    \caption{Two-neutron separation energy residuals, $S_{2n,\textrm{cal}} - S_{2n,\textrm{exp}}$, obtained with \textsc{SkyAx} \cite{Reinhard2021}. The dotted line represents $N=Z$. Solid lines show magic numbers. Upper: SQMC. Lower: SLy4d.}
    \label{fig:S2n_residuals_SQMC_SLy4d}
\end{figure}
\begin{figure}[!htb]
    \centering
    \includegraphics[width=0.9\columnwidth]{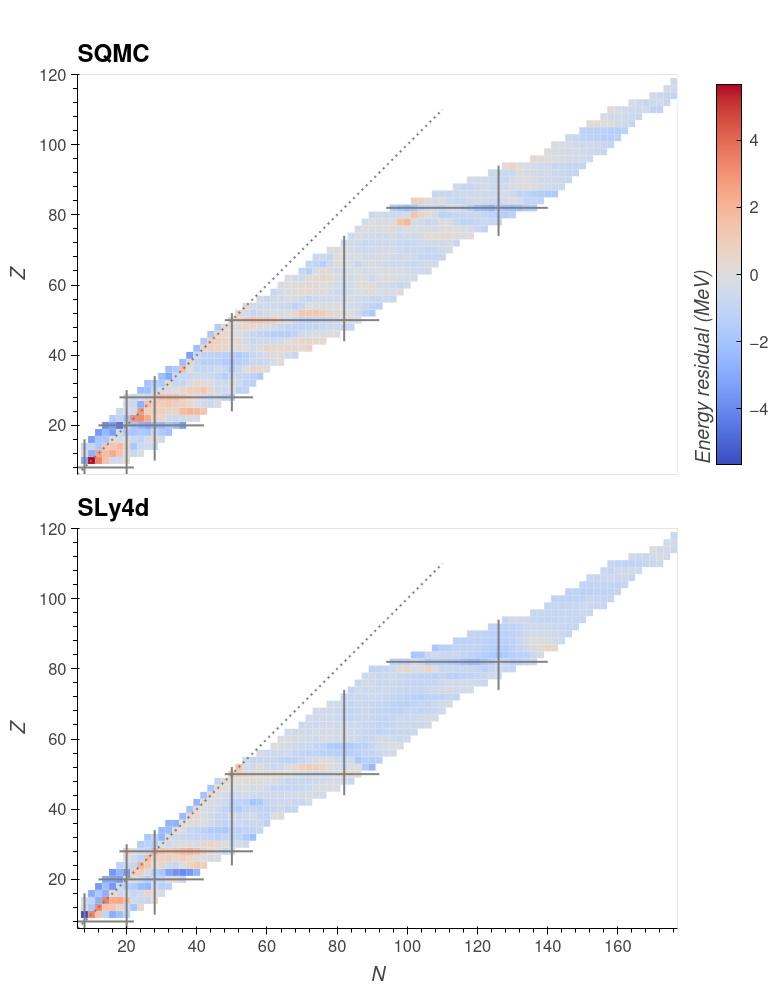}
    \caption{Same as Fig. \ref{fig:S2n_residuals_SQMC_SLy4d} for two-proton separation energy residuals.}
    \label{fig:S2p_residuals_SQMC_SLy4d}
\end{figure}
Looking at two-neutron separation energy residuals in Figure~\ref{fig:S2n_residuals_SQMC_SLy4d}, we see that both functionals show strong bands of red immediately above the $N=82$ shell closure, and blue immediately below. SQMC also displays this pattern across the $N=126$ shell, while it is much less distinct for SLy4d. Across the $N=50$ shell, the bands are much weaker for both functionals and are skewed away from $N=50$ for lighter nuclei.
This pattern is due to the predicted separation energies changing more rapidly across the shell closure from one nucleus to the next along the isotopic chain, than the experimental values which change more smoothly.
\par
The residuals for both functionals show significant variation for nuclei below $N=50$, though the patterns are similar. Most nuclei lying on or above the line of symmetry (i.e. $Z\ge N$) have an underestimated separation energy, while those just below the line of symmetry are overestimated, excluding the region around $^{48}$Ca.
\par
Figure~\ref{fig:S2p_residuals_SQMC_SLy4d} displays two-proton separation energy residuals for SQMC and SLy4d.
Overall, both functionals tend to overestimate the strength of two-proton separation energies. 
SQMC has distinct bands of overestimation (blue) for $Z=20$, $Z=40$, and $Z=82$.
SLy4d, on the other hand, gives a broad region of overestimation around $^{208}$Pb, as well as broad bands for $Z=32-34, 40-42, 54-58$, and a tight band at $Z=22$.
\par
Both functionals overestimate the separation energy for nuclei lying on or above the line of symmetry. 
Just below the line of symmetry there are broad regions of underestimated separation energies around $Z=12$ and $Z=26$.
\par
For both two-neutron and two-proton separation energies, the band structures present in the residuals indicate that there are some shell effects that are not fully captured by either functional.
SQMC is performing with similar accuracy to SLy4d, yielding predictions that are slightly better in some areas of the nuclear chart, and slightly worse in others.

\subsection{Charge radii}
\par
A density distribution, $\rho_q$, can be Fourier transformed to give a `form factor' in momentum-space
\begin{align}
F_q(\bm{k}) = \int \! d\bm{r}\, e^{i \bm{k}\cdot \bm{r} } \rho_q(\bm{r}).
\end{align}
The radially averaged charge form factor is obtained by folding with the intrinsic nucleon form factors~\cite{Bender2003}
\begin{align}
F_c(k) = \sum_{q}\left(F_q G_{E,q} + F_{ls,q} G_{M}\right) \exp\left(\frac{\hbar^2k^2}{8\left\langle\hat{\bm{p}}^2_{cm}\right\rangle}\right)
\end{align}
where 
$F_q$ are form factors for the point nucleon distributions,
$F_{ls,q}$ is proportional to the form factor for the gradient of the spin-orbit density ${\nabla\cdot\bm{J}_q}$ (see Eq.~\ref{eq:spin_orbit_density}),
$G_{E/M}$ are the intrinsic electric and magnetic `Sachs' form factors of nucleons~\cite{Simon1980}, and
the final term is a correction for centre-of-mass motion.
For calculations using \textsc{SkyAx}, $G$ is approximated as a sum of monopoles to order 4~\cite{Simon1980}, and the magnetic term is neglected
\cite{Reinhard2020Mendeley}.
\par
The charge form factor can be Fourier transformed back to a density distribution, from which the mean-square radius is calculated 
\begin{align}
r_{c} = \sqrt{\left\langle r_{c}^2\right\rangle} \textrm{, where }
\left\langle r_{c}^2\right\rangle = \frac{\int \! d\bm{r}\,\bm{r}^2 \rho_c(\bm{r})}{\int \! d\bm{r}\,\rho_c(\bm{r})}.
\end{align}
\par
\begin{figure}[!htb]
    \centering
    \includegraphics[width=0.9\columnwidth]{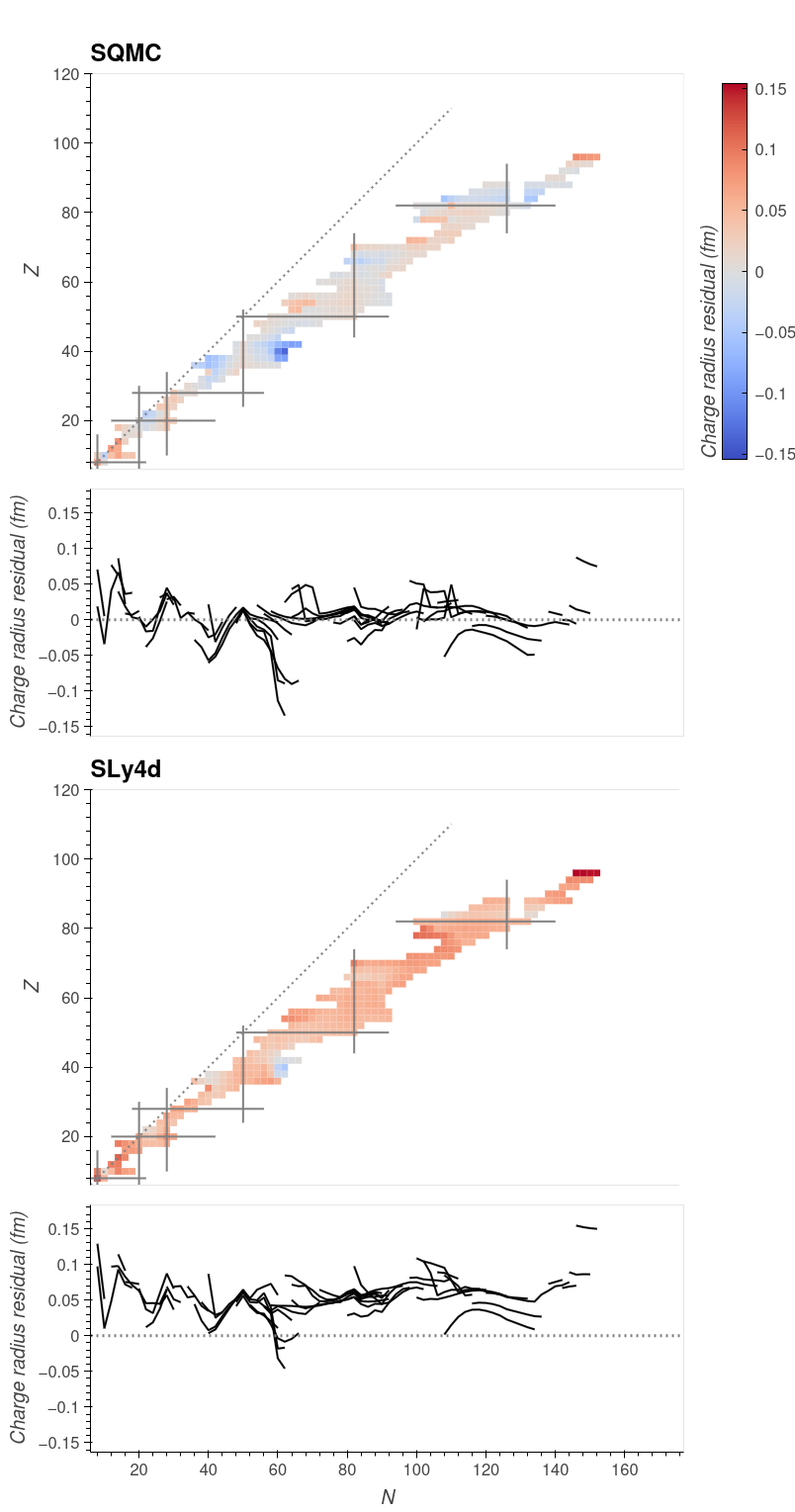}
    \caption{Root-mean-square charge radius residuals, $r_{c,\textrm{cal}} - r_{c,\textrm{exp}}$, obtained with \textsc{SkyAx} \cite{Reinhard2021}. Upper: SQMC. Lower: SLy4d. In the 2\textsuperscript{nd} and 4\textsuperscript{th} panels, each line corresponds to an isotopic chain.}
    \label{fig:Rc_residuals_SQMC_SLy4d}
\end{figure}
Figure~\ref{fig:Rc_residuals_SQMC_SLy4d} shows root-mean-square charge radii, relative to experimental data, for the SQMC and SLy4d functionals. Experimental data is taken from Ref.~\cite{Angeli2013}. Overall, the results from SQMC are markedly closer to experiment than SLy4d, for the majority of the nuclei where experimental data is available (see Tab.~\ref{tab:RMS_all_observables}).
\begin{table}[htb]
   \caption{\label{tab:RMS_all_observables}RMS residuals from experimental values for SQMC and SLy4d. Observables are binding energy $E$, two-neutron separation energy $S_{2n}$, two-proton separation energy $S_{2p}$, and RMS charge radii $r_c$.}
   \centering
   \begin{tabular}{l|l|l|l}
   Observable     & SLy4d & SQMC & No. nuclei \\
   \hline
   $E$ (MeV)      & 3.40  & 2.79 & 865 \\
   $S_{2n}$ (MeV) & 0.982 & 0.959 & 809 \\
   $S_{2p}$ (MeV) & 1.10  & 1.02 & 780 \\
   $r_{c}$ (fm)   & 0.0598 & 0.0269 & 345 \\
    \end{tabular}
\end{table}
Additionally, the distribution of residuals for SQMC is centred around zero, while SLy4d has a tendency to overestimate the charge radii. 
{ However, a direct statistical comparison should be interpreted with some caution because the SLy4d parameters were adjusted using a different prescription for the charge radius corrections \cite{Chabanat1998} than the one implemented in \textsc{SkyAx}.}
\par
Both functionals underestimate radii for nuclei around $^{102}$Zr ($N=62,Z=40$), SQMC to a larger degree than SLy4d. The experimental radii for Zr isotopes feature a rapid increase between $^{99}$Zr and $^{100}$Zr as the nuclei become substantially deformed~\cite{Raman2001}. The calculated results are unable to capture this behaviour in either the charge radii or deformation, as will be discussed below.
\par
Looking to the top end of the nuclear chart, both functionals overestimate the radii of Cm ($Z=96$) isotopes. This is due to a turning point in the experimental values, with Cm isotopes having smaller radii than the corresponding Pu ($Z=94$) and Am ($Z=95$) isotopes below them. The calculated results do not replicate this, continuing on an upward trend.
\par
\begin{figure}[!htb]
    \centering
    \includegraphics[width=0.8\columnwidth]{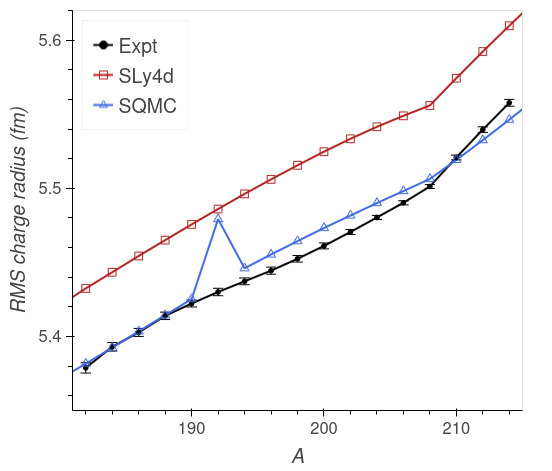}
    \caption{Root-mean-square charge radii, $r_{c}$, for the Pb isotope chain, obtained with \textsc{SkyAx} \cite{Reinhard2021}. Experimental data from Ref.~\cite{Angeli2013}.}
    \label{fig:Pb_radii}
\end{figure}
Figure~\ref{fig:Pb_radii} shows RMS charge radii as calculated by SQMC and SLy4d, compared to experimental values which display a characteristic `kink' at $^{208}$Pb.
{SQMC closely matches the experimental values for all isotopes except for $^{192}$Pb, which SQMC predicts to have an oblate deformed ground state. }
\par
SQMC yields a less pronounced isotope kink than either experiment or SLy4d.
It is interesting to note that the reproduction of this kink was the initial motivation for allowing $W_0\ne W_0'$ in Ref. \cite{Reinhard1995}. However, leaving $W'_0/W_0$ free in the fit led to the SkI4 parameterisation with  $W_0\simeq-W_0'>0$, in contradiction with QMC and UNEDF1 which predict $W_0'>W_0>0$. The kink could also be reproduced in \cite{Reinhard1995} by keeping $W'_0/W_0=1$ and lowering the effective mass. This might be the reason why the kink is observed in SLy4d which has a low isoscalar effective mass at saturation density  $m^*/m\simeq0.66$ (compare with $m^*/m\simeq0.79$ for SQMC). 

\subsection{Quadrupole deformation}\label{sec:SystematicDeformation}
The axial quadrupole moment, and dimensionless $\beta_2$, are derived from a density distribution as
\begin{align}
Q_{20} \equiv Q_2 &= \int \! d\bm{r}\, r^2 Y_{20} \rho(\bm{r})\\
\beta_2 &= \frac{4\pi}{3 A R_0^2} Q_2
\end{align}
where $r = |\bm{r}|$,  $R_0 = 1.2\times A^{1/3}$~fm is a reference radius, and $Y_{20}$ is the spherical harmonic.
\par
\begin{figure}[!htb]
    \centering
    \includegraphics[width=0.9\columnwidth]{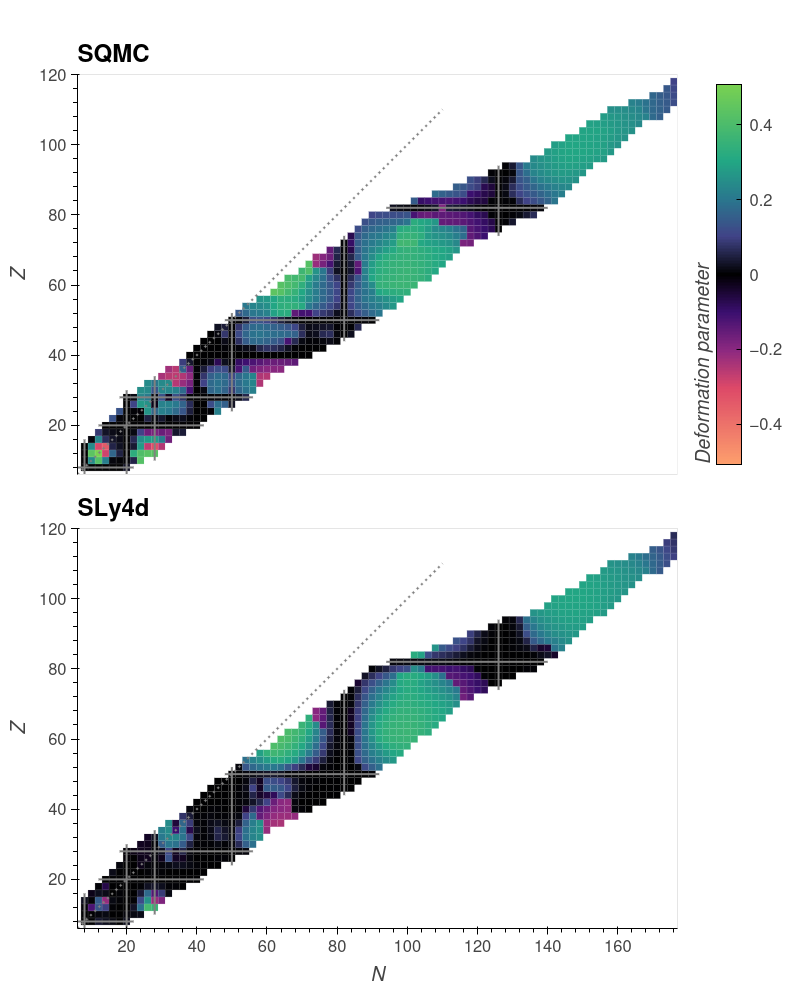}
    \caption{Quadrupole deformation parameter, $\beta_{2,\textrm{cal}}$, obtained with \textsc{SkyAx} \cite{Reinhard2021}. Upper: SQMC. Lower: SLy4d.}
    \label{fig:beta2_SQMC_SLy4d}
\end{figure}
\par
Systematic calculations of the quadrupole deformation parameter are shown in Figure~\ref{fig:beta2_SQMC_SLy4d}. 
At a high level, SQMC and SLy4d produce similar patterns of ground state deformation. 
{
Large islands of prolate deformation are separated by bands of spherical nuclei running along the magic numbers, and small patches of oblate deformation.
The largest deformations occur for the mid-mass nuclei between Sn ($Z=50$) and Pb ($Z=82$), as well as some small regions of very strong prolate deformation for Mg ($Z=12$) and oblate deformation for Si ($Z=14$) predicted by SQMC.
\par
SQMC predicts the semi-magic Zr ($Z=40$) isotopes to be spherical, with prolate deformation for the isotopes above, and a mixture of oblate and prolate regions immediately below $Z=40$.
SLy4d does not feature a strong spherical band, with several Zr isotopes predicted to be oblate deformed.
}
Overall, SQMC and SLy4d predictions of deformation parameters agree for $Z\gtrsim40$ but significant differences are found for some lighter nuclei.
\par
\begin{figure}[!htb]
    \centering
    \includegraphics[width=0.9\columnwidth]{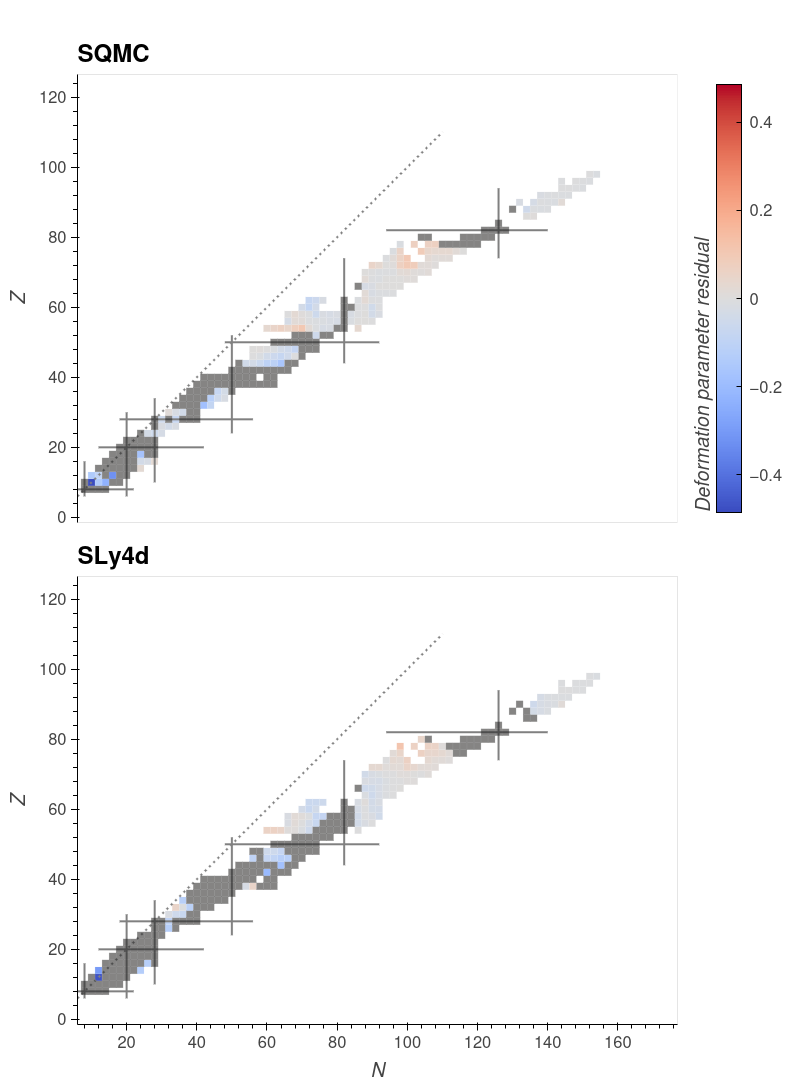}
    \caption{Quadrupole deformation parameter residuals, $|\beta_{2,\textrm{cal}}| - |\beta_{\textrm{exp}}|$, obtained with \textsc{SkyAx} \cite{Reinhard2021}. Upper: SQMC. Lower: SLy4d. Note: dark grey indicates nuclei predicted to be nearly spherical with $|\beta_{2,\textrm{cal}}|<0.1$.}
    \label{fig:beta2_residual}
\end{figure}
Figure~\ref{fig:beta2_residual} shows the calculated deformation parameter, $\beta_2$, relative to the $\beta$ parameter derived from experimental data of the reduced electric quadrupole transition probability from the ground state to the first $2^+$ state, $B(E2)\!\uparrow$~\cite{Raman2001}. In this comparison we filter to only well deformed nuclei where we can reasonably expect the experimental $\beta$ value to be caused primarily by static deformation, i.e. where $|\beta_{2,\textrm{cal}}|\ge0.1$. It should also be noted that the experimental values are strictly positive. The two functionals predict values of $\beta_2$ in line with experiment for most nuclei above $Z=50$, with some overestimation for nuclei around $^{178}$Os ($Z=76, N=102$). For lighter nuclei, below $Z=50$, both functionals tend to predict weaker deformation than reality.

\section{Isovector dependence of the spin-orbit functional}
\label{sec:IsovectorSO}
The spin-orbit interaction is required for the correct prediction of all magic numbers above 20. 
It is also a crucial dissipation mechanism in heavy-ion collisions~\cite{umar1986},
and it is expected to be significant in determining the location of the so-called ``superheavy island of stability''~\cite{bender1999}.
Its microscopic origin is still debated \cite{ebran2016,Chanfray2020,Ding2026}.
\par
Non-relativistic approaches, such as the Skyrme EDF, require experimental data to determine the strength of the spin-orbit interaction.
Often these data consist of single-particle energy level splittings.
While this approach is efficient to constrain a single spin-orbit parameter ($W_0=W_0'$), the isovector dependence of the spin-orbit interaction (controlled by the introduction of a second spin-orbit parameter $W_0'\ne W_0$ as in Eq.~\ref{eq:HsoSkyrme}) is more difficult to constrain and often neglected.
Table~\ref{tab:W0'/W0} compares the ratio of $W_{0}'$ to $W_{0}$ for different approaches.
\par
\begin{table}[htb]
   \caption{\label{tab:W0'/W0}Ratio of the spin-orbit coupling constants, illustrating the isovector dependence of spin-orbit energy density functionals.}
   \centering
   \begin{tabular}{l|l}
   \hline\\[-1.0em]
   Model & $W_{0}'/W_{0}$ \\
   \hline
   Hartree & 0 \\
   Hartree-Fock (Skyrme force) & 1 \\
   UNEDF1 (Skyrme functional) & 1.86 \\
   SQMC & 1.78 \\
   \hline
	\end{tabular}
\end{table}
\par
If we only include Hartree terms, the zero-range, two-body spin-orbit force of the Skyrme interaction produces a functional with only an isoscalar term and $W_{0}'=0$.
Inclusion of the Fock (exchange) term introduces an isovector component, but fixes $W_{0}' = W_{0}$ so the functional still has only one spin-orbit parameter. This is the form used by the majority of  Skyrme parameterisations, in particular those based on an underlying zero-range Skyrme interaction.
\par
It is possible to introduce a second independent spin-orbit parameter to the Skyrme EDF by relaxing its connection to the two-body Skyrme force \cite{Reinhard1995}.
This approach as been used in the UNEDF1 parameterisation~\cite{Kortelainen2012}. 
An extensive fit to medium and heavy mass nuclei yielded a significantly stronger isovector dependence for UNEDF1 than standard Skyrme forces (see Tab.~\ref{tab:W0'/W0}).
\par
As outlined in Sec~\ref{sec:QMCEDF}, the QMC EDF is obtained as the non-relativistic limit of a relativistic treatment of quarks and mesons.
It emerges as a combination of the variation of the vector meson fields, $\omega$ and $\rho$, across the finite size of the nucleon, and the purely relativistic effect of Thomas precession when changing frames of reference. As the resulting functional (Eq.~\ref{eq:HsoQMC}) emerges naturally from the relativistic treatment of QMC, the spin-orbit term has no additional free parameters, only the three coupling constants and $\sigma$ meson mass fixed by the central terms.
\par
The SQMC parameterisation fully reproduces the isovector dependence of the QMC spin-orbit functional. Using the parameter values described in Sec.~\ref{sec:QMCconstants}, the QMC and SQMC functionals possess a strong spin-orbit isovector dependence.
As listed in Tab.~\ref{tab:W0'/W0}, the ratio of parameters for SQMC is substantially larger than one, and remarkably close to the fitted value of UNEDF1.
Additionally, Ref.~\cite{Stone2016}, which fixes the free parameters of the QMC model through a fit to experimental data, obtained spin-orbit parameters very close to those of SQMC, with a ratio of $W_{0}'/W_{0}=1.83$.
\par
 It should be noted that the version of QMC used to derive SQMC ignores several effects that have been included in more recent versions of QMC, such as pions \cite{Guichon2018} and tensor terms \cite{Martinez2020}. In particular, no finite range exchange mechanisms are accounted for. These effects could naturally have a quantitative impact on $W_{0}'/W_{0}$. The evaluation of these effects on the isovector dependence of the spin-orbit term of the functional would require further extensions of SQMC, e.g., with additional tensor terms \cite{lesinski2007}, that are beyond the scope of this work. 

The following discussion will isolate the impact of only the isovector dependence of the spin-orbit term by comparing the SQMC parameterisation (with $W_{0}'/W_{0}=1.78$) against a baseline parameterisation with a single spin-orbit parameter ($W_{0}'=W_{0}\equiv\tilde{W}_0=104.3872$~MeV~fm$^5$) but that is otherwise identical to SQMC.
This new value for the spin-orbit parameter is obtained by equating $\Ham_{\textrm{SO}}^{\textrm{Skyrme}}$ in Eq.~\ref{eq:HsoSkyrme} for symmetric systems ($\rho_q=\rho/2$ and $\bm{J}_q=\bm{J}/2$), leading to $\tilde{W}_0=\frac{2}{3}{W_0}+\frac{1}{3}{W_0'}$.
\par
\begin{figure}[hbt]
   \centering
   \includegraphics[width=0.8\columnwidth]{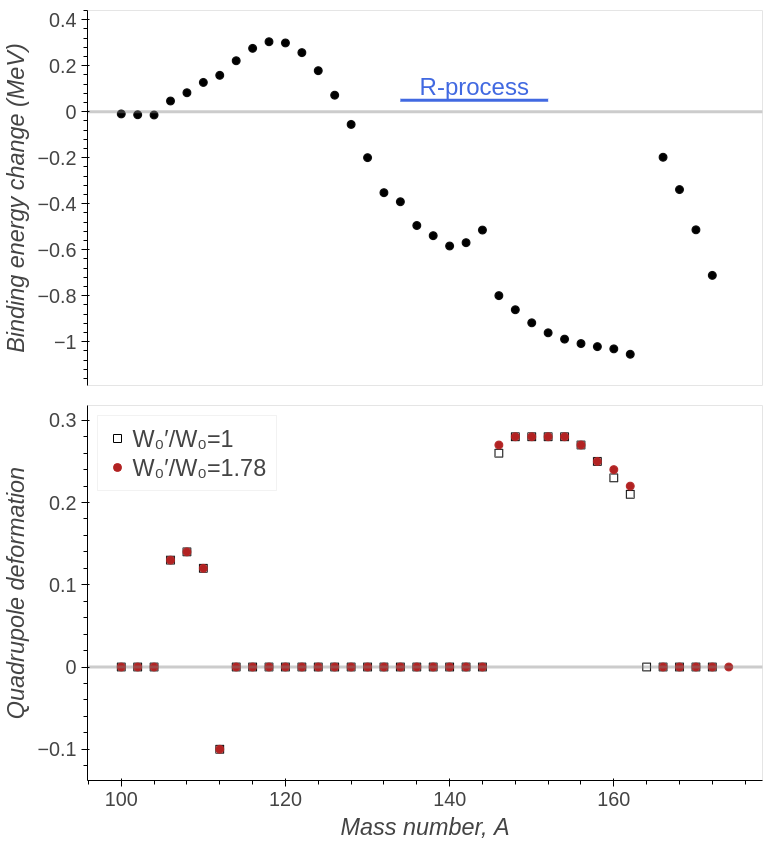}
   \caption{\label{fig:EBchange_Sn}Upper: Binding energy difference between SQMC with ${W'_{0}/W_{0}=1.78}$ and SQMC with ${W'_{0}/W_{0}=1}$ in tin isotopes. Lower: Quadrupole deformation, $\beta_{2}$, predicted by the two SQMC parameterisations. Results obtained with \textsc{hfbtho} \cite{Stoitsov2013}.} 
\end{figure}
\begin{figure}[hbt]
   \centering
   \includegraphics[width=0.8\columnwidth]{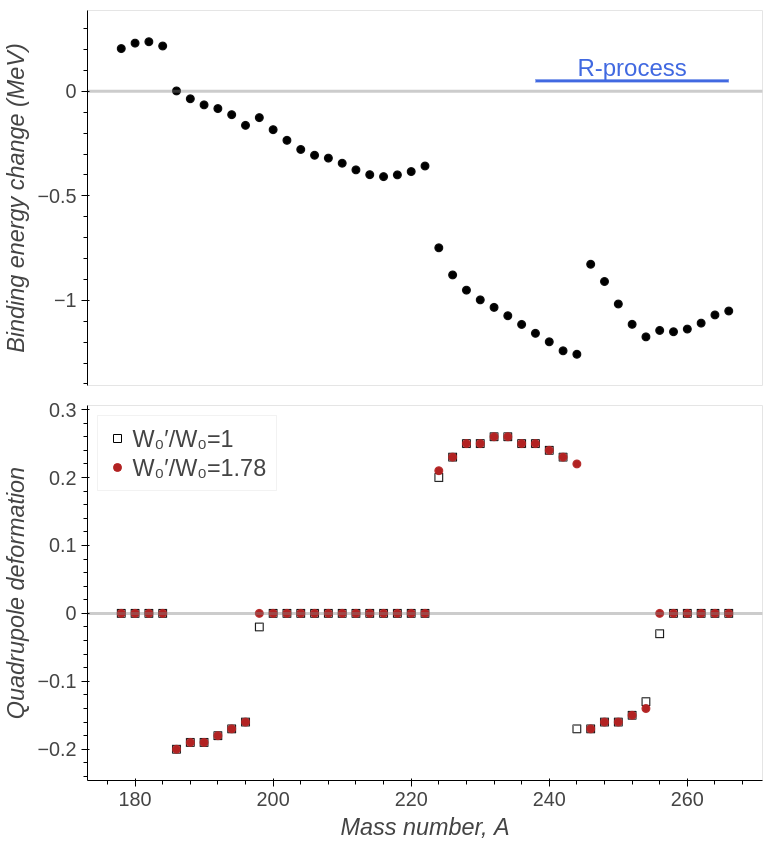}
   \caption{\label{fig:EBchange_Pb} Same as Fig.~\ref{fig:EBchange_Sn} for lead isotopes.}
\end{figure}
\begin{figure}[hbt]
   \centering
   \includegraphics[width=0.8\columnwidth]{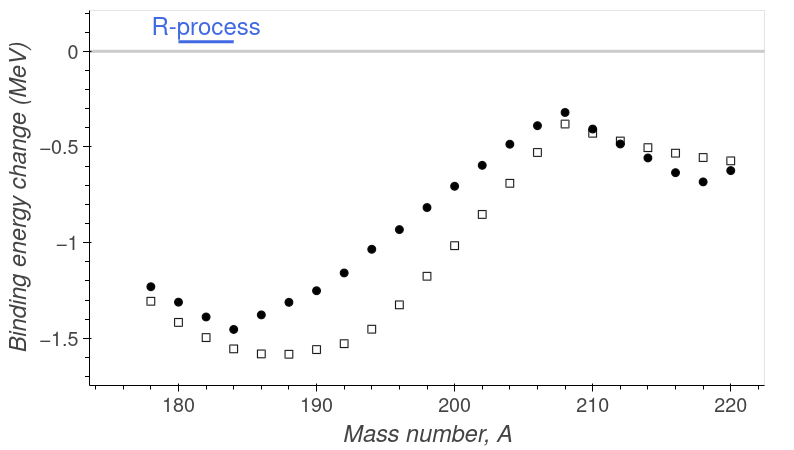}
   \caption{\label{fig:EBchange_N126}$N=126$ isotonic chain: Binding energy difference between SQMC with ${W'_{0}/W_{0}=1.78}$ and SQMC with ${W'_{0}/W_{0}=1}$. Open squares are the equivalent quantity calculated with $m_\sigma=600$~MeV. Results obtained with \textsc{hfbtho} \cite{Stoitsov2013}.}
\end{figure}
\textsc{hfbtho} calculations were performed for all bound nuclei of the Sn and Pb isotopic chains and the $N=126$ isotonic chain, and are reported in Figures~\ref{fig:EBchange_Sn}, \ref{fig:EBchange_Pb} and \ref{fig:EBchange_N126}, respectively.
The \textsc{hfbtho} code is chosen for this investigation because the Hartree-Fock-Bogoliubov (HFB) treatment of pairing remains robust for weakly bound nuclei, unlike the BCS pairing approximation employed by \textsc{SkyAx}.
\par
The upper panel of Figure~\ref{fig:EBchange_Sn} shows the difference in ground state binding energy between SQMC with ${W'_{0}/W_{0}=1.78}$ and SQMC with ${W'_{0}/W_{0}=1}$, while the lower panel shows the predicted quadrupole deformation. 
Only  nuclei with a positive two-neutron separation energy have been included.
The two functionals give slightly different predictions of the location of the two-neutron dripline, with $^{174}$Sn predicted to be weakly bound for SQMC with ${W'_{0}/W_{0}=1.78}$, and slightly unbound for $W'_{0}/W_{0}=1$. Additionally, $^{164}$Sn is slightly unbound for $W'_{0}/W_{0}=1.78$.
Binding energy changes are only reported for cases where both functionals predict bound nuclei.
\par
The two functionals give very similar results for the quadrupole deformation of Sn nuclei. 
Most cases are calculated to be spherical, with prolate regions for $^{106 - 110}$Sn and $^{146 - 162}$Sn, and the single oblate case
of $^{112}$Sn.
\par
The lighter nuclei in the region of $^{106 - 126}$Sn show a positive energy change of up to 300~keV, indicating that the functional with $W'_{0}/W_{0}=1$ is more strongly bound for these nuclei.
This reverses for all nuclei above $^{128}$Sn, with ${W'_{0}/W_{0}=1}$ now less bound than ${W'_{0}/W_{0}=1.78}$.
\par
A sharp discontinuity is observed in the trend of energy change for $^{146 - 162}$Sn, corresponding to the region of prolate deformation. The energy difference between the two functionals appears to be enhanced by the deformation, reaching a change of up to 1~MeV. 
\par
Figure~\ref{fig:EBchange_Pb} shows the same quantities  for the Pb isotopic chain.
The functional with ${W'_{0}/W_{0}=1.78}$ predicts less bound nuclei for $^{178 - 186}$Pb, and more strongly bound heavier lead isotopes.
The energy change between the two functionals reaches a maximum of 1.3~MeV, again displaying discontinuities along the isotopic chain that appear to correspond with changes in deformation.
\par
The lower panel of figure~\ref{fig:EBchange_Pb}  reports multiple regions of oblate and prolate deformations.
The SQMC functional with $W'_{0}/W_{0}=1.78$ predicts that $^{186 - 196}$Pb and $^{246 - 254}$Pb have oblate minima, while $^{224 - 244}$Pb are prolate, and the remainder isotopes are spherical.
The $W'_{0}/W_{0}=1$ case generally agrees, except for $^{244}$Pb which is oblate for ${W'_{0}/W_{0}=1}$ and prolate for $W'_{0}/W_{0}=1.78$.
 Note that the predicted oblate deformation of $^{186 - 196}$Pb is not in agreement with experiments that assign spherical ground-states to these nuclei \cite{julin2016}. Beyond mean-field calculations are necessary to reproduce the shape coexistence in this region \cite{rodriguez-guzman2004}.
\par
Figure~\ref{fig:EBchange_N126} shows the binding energy difference between the SQMC functional with ${W'_{0}/W_{0}=1.78}$ and ${W'_{0}/W_{0}=1}$ for the $N=126$ isotonic chain, equivalent to the upper panels of Figs.~\ref{fig:EBchange_Sn} and~\ref{fig:EBchange_Pb}. All nuclei are predicted to be spherical by both parameterisations, so the lower panel is omitted.
The SQMC functional with ${W'_{0}/W_{0}=1.78}$ predicts more strongly bound nuclei for all $N=126$ cases.
The smallest difference between the two functionals is seen in $^{208}$Pb, while the largest change of almost 1.5~MeV occurs for $^{184}$Ce.
\par
The open squares shown in Fig.~\ref{fig:EBchange_N126} display the  binding energy difference induced by the isovector dependence of the spin orbit term, using SQMC functionals evaluated with $m_\sigma=600$~MeV, rather than 700~MeV.
While the isovector channel, through $G_\rho$, is more sensitive to the choice of $m_\sigma$ (see Fig.~\ref{fig:coupling_constants}), leading to a slightly larger magnitude of the binding energy difference, its pattern is similar, indicating that these results do not significantly depend on $m_\sigma$. 

\par
Overall, as we move away from symmetry (i.e. to the right in Figs.~\ref{fig:EBchange_Sn} and \ref{fig:EBchange_Pb}, and to the left in Fig.~\ref{fig:EBchange_N126}) we see the energy change growing rapidly.
Inclusion of the spin-orbit isovector dependence of SQMC tends to yield more strongly bound nuclei, particularly for more neutron rich systems.
\par
While the energy changes induced by the isovector dependence of the spin-orbit interaction are small, it is expected that a change of as little as 500~keV could have a significant impact on r-process abundances~\cite{Mumpower2016}.
\par
The r-process is expected to follow a path of constant single-neutron separation energies with $S_{1n}\approx 2-3$~MeV~\cite{Langanke2001,Kratz1993}.
As this work is restricted to even-even nuclei, we use the approximation of Ref.~\cite{Erler2012} for the determination of single-neutron separation energies\footnote{Note that here we use the convention where $S_{1n}$ corresponds to the energy required to remove a single neutron, rather than the energy liberated.}
\begin{align}
   S_{1n}(Z, N) &= E(Z, N - 1) - E(Z, N) && \\
   &\approx -\lambda_n(N) - \Delta_n(N) &&\textrm{, for odd N nuclei,}
\end{align}
where $\lambda_n$ is the Fermi energy and $\Delta_n$ is the pairing gap for the neutrons of the nucleus with $N$ neutrons.
The Fermi energy and pairing gap of the odd nucleus is approximated as the average of the neighbouring even-even nuclei~\cite{Erler2012}, giving
$$   S_{1n}(N) \approx -\frac{1}{2}\left( \lambda_n(N-1) + \lambda_n(N+1) + \Delta_n(N-1) + \Delta_n(N+1) \right) $$
    for odd N, and 
$$   S_{1n}(N) \approx E(N-2) - S_{1n}(N-1) - E(N) $$
for even $N$. 

Figures~\ref{fig:EBchange_Sn}, \ref{fig:EBchange_Pb} and \ref{fig:EBchange_N126} include annotations for the r-process regions of Sn, Pb and $N=126$, respectively. The labels correspond to the region where even-even nuclei have $1.75 < S_{1n} < 3.25$ MeV\footnote{This window of $S_{1n}$ is broader  than the $2-3$~MeV  previously mentioned. This larger window allows for uncertainties in the approximations used to compute separation energies.} as predicted by the SQMC functional with $W'_{0}/W_{0}=1.78$. This procedure yields $^{134 - 152}$Sn, $^{238 - 266}$Pb and $^{180 - 184}Z_{126}$ as the predicted regions relevant for the r-process.
\par
In the regions of the r-process, the energy changes caused by the isovector dependence of the spin-orbit interaction range from 0.5~to~1.5~MeV. While the differences are small in absolute terms, they may be sufficient to impact the outcomes of r-process nucleosynthesis~\cite{Mumpower2016}.
It is therefore important to properly account for the isovector properties of the spin-orbit functional when studying exotic nuclei.
 Note that other terms in the functional could have larger uncertainties than the isovector dependence of the spin-orbit. In particular, uncertainties on the symmetry energy (see, e.g., \cite{lattimer2023}) could propagate to uncertainties on the drip line position. The impact of the isovector spin-orbit is still significant enough to be noted, and should be considered in addition to improvement of other terms in the functional. Note also that in \cite{Martinez2020}, the QMC parameters have been fitted with an associated uncertainty that propagates to the symmetry energy, giving 29(1) MeV, compatible with the SQMC value in Tab~\ref{tab:INMproperties}.
\par
\begin{figure}[hbt]
   \centering
   \includegraphics[width=\columnwidth]{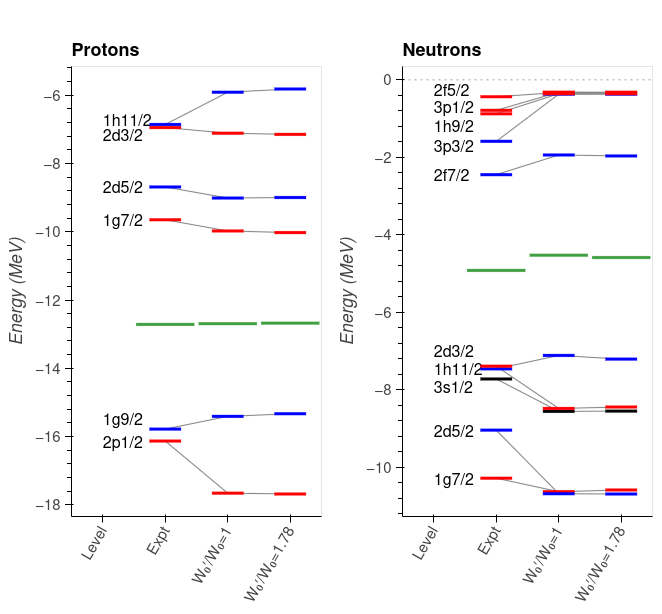}
   \caption{\label{fig:spel_Sn132}Single-particle energy spectra of $^{132}$Sn for experimental data~\cite{Grawe2007}, and \textsc{hfbtho} calculations without Lipkin-Nogami projection, of SQMC with ${W'_{0}/W_{0}=1.78}$ and SQMC with ${W'_{0}/W_{0}=1}$. The green bands estimate the Fermi level.} 
\end{figure}
{
Figure~\ref{fig:spel_Sn132} shows the single-particle energy spectrum of $^{132}$Sn, calculated using \textsc{hfbtho} without Lipkin-Nogami corrections.
Calculations are compared to experimental data collated in Ref.~\cite{Grawe2007}.
Red and blue levels indicate $j=l-\frac{1}{2}$ and $j=l+\frac{1}{2}$ spin-orbit partners, respectively.
Changes to the spectrum induced by the stronger isovector dependence of the spin-orbit term are very small.
For protons, $j=l-\frac{1}{2}$ partners tend to move down slightly, while $j=l+\frac{1}{2}$ move up.
For neutrons, the opposite behaviour is observed.
As a result, the isovector spin-orbit term tends to slightly reduce (resp. increase) the spin-orbit splitting for protons (neutrons). 
However, the dependence of the single particle energy levels with the isovector spin-orbit term of the functional is observed to be very weak, which is in line with the findings of Ref.~\cite{Kortelainen2008}.
}

\section{Conclusions}
\label{sec:Conclusion}

In this work we reviewed the feasibility of using a Skyrme functional inspired by the QMC model for the study of nuclear structure. This was done through a systematic study of the nuclear chart, followed by focussed investigations of specific isovector terms of the functional.
\par
SQMC delivers a similar level of accuracy when predicting the properties of finite nuclei across the nuclear chart, as compared to traditional Skyrme functionals such as SLy4d.
The systematic studies of ground state binding energies, two-particle separation energies, charge radii, and quadrupole deformations all show acceptable performance by SQMC.
For example, SQMC yields a RMS residual of 2.79~MeV for binding energies of all even-even nuclei with a known mass, with proton and neutron numbers greater than 7, which is of similar order to standard Skyrme approaches. 
{Recall that the physics-driven inputs of SQMC are inherited from the QMC model, whose four free parameters are fixed using infinite nuclear matter pseudodata. The mapping of the QMC functional onto a Skyrme form additionally involves conventional EDF modelling choices, as discussed in Sec.~\ref{sec:SQMCfitting}. In contrast, traditional phenomenological Skyrme functionals generally involve many more adjustable parameters (\(\gtrsim 10\)) and are typically fitted to a combination of nuclear matter properties and experimental data on finite nuclei.}
Mapping the QMC functional to a Skyrme one also fixes (in a phenomenological way) the issue of the large incompressibility of the original QMC model (a feature considerably improved in more recent versions of QMC including pions \cite{Guichon2018}). Although this leads to a clear departure from the original QMC prediction of the central force, other terms of the functional, such as the spin orbit term and its isovector dependence, are preserved in SQMC.
\par
The spin-orbit term of SQMC functional, including its isovector dependence, arises naturally from the relativistic QMC model, with no additional free-parameters.
The spin-orbit isovector dependence of SQMC is much stronger than what is used by most standard Skyrme parameterisations, and very similar to that of UNEDF1.
Changing the isovector dependence of the spin-orbit functional, from $W_{0}'/W_{0}=1.78$ as predicted by the QMC model, to $W_{0}'/W_{0}=1$ as used by most standard Skyrme parameterisations, can induce binding energy changes of 1~-~1.5~MeV for very neutron rich isotopes of Sn and Pb, and $N=126$ isotones. While this is a small change, it may be significant when modelling certain exotic systems, such as nuclei relevant for the astrophysical r-process.

\section*{Acknowledgments}
We are grateful to A.W. Thomas for useful discussions on the QMC model.
This work was supported by computational resources provided by the Australian Government through the National Computational Infrastructure (NCI) under the ANU Merit Allocation Scheme.

\appendix 
\section*{APPENDICES}
\section{SQMC in Hartree-Fock solvers}\label{sec:codes}

\par

\subsection{\textmd{\textsc{SkyAx}}}\label{sec:SkyAx}
\textsc{SkyAx}~\cite{Reinhard2021} is a Skyrme-Hartree-Fock code for the calculation of static nuclear properties and deformation energy surfaces. It allows for only axial deformation, and includes pairing under the BCS approximation.
In the development of \textsc{SkyAx}, particular consideration was given to computational speed.
\par
A BCS stabilising pairing energy, $E_{\textrm{stab}}$, is used as an alternative to Lipkin-Nogami particle number projection~\cite{Erler2008}. It is a numerical parameter that directly modifies the pairing gaps, $\Delta_{\alpha}$ from Eq.~\ref{eq:pairing_gap_k}, such that
\begin{align}
    \Delta_{\alpha} \rightarrow \Delta_{\alpha} \left(1+ \frac{E_{\textrm{stab}}^2}{E_{\textrm{pair},q_\alpha}^2}\right),
\end{align}
where $E_{\textrm{pair},q_\alpha}$ is the pairing energy for the nucleon type of state $\alpha$~\cite{Reinhard2021}.
\textsc{SkyAx} calculations reported in this work use a pairing energy functional defined as 
\begin{align}
    E_{\textrm{pair}}=\frac{1}{4}\sum_{q=n,p} V_{{0},q} \int d\bm{r} \left|\xi_q\right|^2\left[1-\frac{\rho}{\rho_{0,\textrm{pair}}}\right] \label{eq:Skyax_pairing}
\end{align}
where $\xi_q$ is the pairing density, and the choice of parameter $\rho_{0,\textrm{pair}}$ controls the contributions of volume and surface pairing.
This is equivalent to a density-dependent delta interaction (DDDI), as described in Eq.~\ref{eq:pairing_int}, up to a constant factor~\cite{Reinhard2021}.
\par
As discussed in Sec.~\ref{sec:pairing}, the pairing parameters are adjusted to reproduce the neutron pairing gap of $^{120}$Sn, with $\Delta_n$ defined by Eq.~\ref{eq:pairing_gap_uv}~\cite{Reinhard2021}. This results in a pairing strength parameter of ${V_0}=853.65$~MeV for SQMC, for both neutrons and protons, using ${E_{\textrm{stab}}=-0.3}$~MeV.
In addition, \textsc{SkyAx} uses a soft cutoff in pairing space (see Sec.~2.4.3 of Ref.~\cite{Reinhard2021} for details). The default cutoff parameter of $\eta_{\textrm{cut}} = 1.65$ has been used in this work.

\subsection{\textmd{\textsc{hfbtho} }}\label{sec:HFBTHO}

\textsc{hfbtho}~\cite{Marevic2022} evaluates the Hartree-Fock equations within a transformed harmonic oscillator basis. Like \textsc{SkyAx}, \textsc{hfbtho} assumes axial symmetry.
Note that in this work, `\textsc{hfbtho}' will always refer to the `v2.00d' version of the code \cite{Stoitsov2013}.
\par
A key distinction of \textsc{hfbtho} is that pairing is included through the full Hartree-Fock-Bogoliubov treatment, rather than the BCS approximation.
BCS can encounter issues with unbound states for nuclei close to the driplines (leading to a so-called `gas in the box'). HFB doesn't share this problem, remaining well defined for weakly bound nuclei~\cite{Bender2003}.
\par
Using HFB pairing with Lipkin-Nogami particle number projection, the DDDI pairing strength (see Eq.~\ref{eq:pairing_int}) for SQMC is adjusted to ${V_{0}=-239.80}$~MeV to reproduce the average neutron pairing gap of $^{120}$Sn, with $\Delta_n$ defined by Eq.~\ref{eq:pairing_gap_v2}~\cite{Stoitsov2005}, 
and a pairing cutoff of $E_{\textrm{cut}}=60$~MeV.
Note that the pairing strength value used in \textsc{hfbtho} is a different sign and magnitude as compared to the value for \textsc{SkyAx} due to the different conventions used by Eqs.~\ref{eq:pairing_int}~and~\ref{eq:Skyax_pairing}.
\par
For the tin isotopic chain, the RMS residual is 1.78 MeV with \textsc{SkyAx} and 1.31 MeV with \textsc{hfbtho}, indicating an improvement with the full HFB treatment. For the lead isotopic chain, however, the RMS residuals are slightly worse with \textsc{hfbtho} (1.92 MeV with \textsc{SkyAx} and 2.09 with \textsc{hfbtho}).

\section{$\Ham_{0+3}$ fit sensitivity}\label{sec:fit_sensitivity}
\begin{table}[htb]
   \caption{\label{tab:fit_sensitivity}Sensitivity of SQMC to changes in choice of $\alpha$ and $\rho$ fit region (fm$^{-3}$). Properties of symmetric nuclear matter: saturation density $\rho_{0}$ (fm$^{-3}$), energy per nucleon at saturation $e_{\infty}$ (MeV), and incompressibility $K_{\infty}$ (MeV). The residual $\Delta E (E_{\textrm{cal}} - E_{\textrm{exp}})$ (MeV) for $^{208}$Pb is calculated with \textsc{hfbtho}. The first line corresponds to SQMC, the second line shows results with an increase of $\alpha$, and the third (resp. fourth) is obtained with a larger (smaller) density fit region parametrised as $\rho_0\pm\Delta\rho$ with $\rho_0=0.16$~fm$^{-3}$ (see text).}
   \centering
   \begin{tabular}{l|l|l|l|l|l}
   \hline\\[-1.0em]
 $\alpha$ & $\Delta\rho$ & $\rho_{0}$ & $e_{\infty}$ & $K_{\infty}$ & $\Delta E(^{208}$Pb)\\
   \hline
 1/6 & 0.04  & 0.159 & -15.83 & 218.7 & -3.04 \\
 1/3 & 0.04  & 0.159 & -15.86 & 247.4 & 19.36 \\
 1/6 & 0.1 & 0.148 & -15.67 & 213.4 & -20.20 \\
 1/6 & 0.02 & 0.161 & -15.93 & 220.2 & -12.96 \\
   \hline
	\end{tabular}
\end{table}
Table~\ref{tab:fit_sensitivity} shows how the properties of SQMC are changed by modifications to the choice of $\alpha$ and the fit region used to determine the parameters ($t_0$, $x_0$, $t_3$, and $x_3$) of the volume term of the EDF, $\Ham_{0+3}$.
The density ranges used in the fits are determined as $\rho_0\pm\Delta\rho$ with $\rho_0=0.16$~fm$^{-3}$. 
Increasing $\alpha$ serves to increase the incompressibility of nuclear matter. The nuclear matter values remain within generally acceptable bounds~\cite{Dutra2012}, however the predictions of the binding energy of $^{208}$Pb are substantially degraded, as compared to the original SQMC parameterisation.

\begin{figure}[hbt]
   \centering
   \includegraphics[width=\columnwidth]{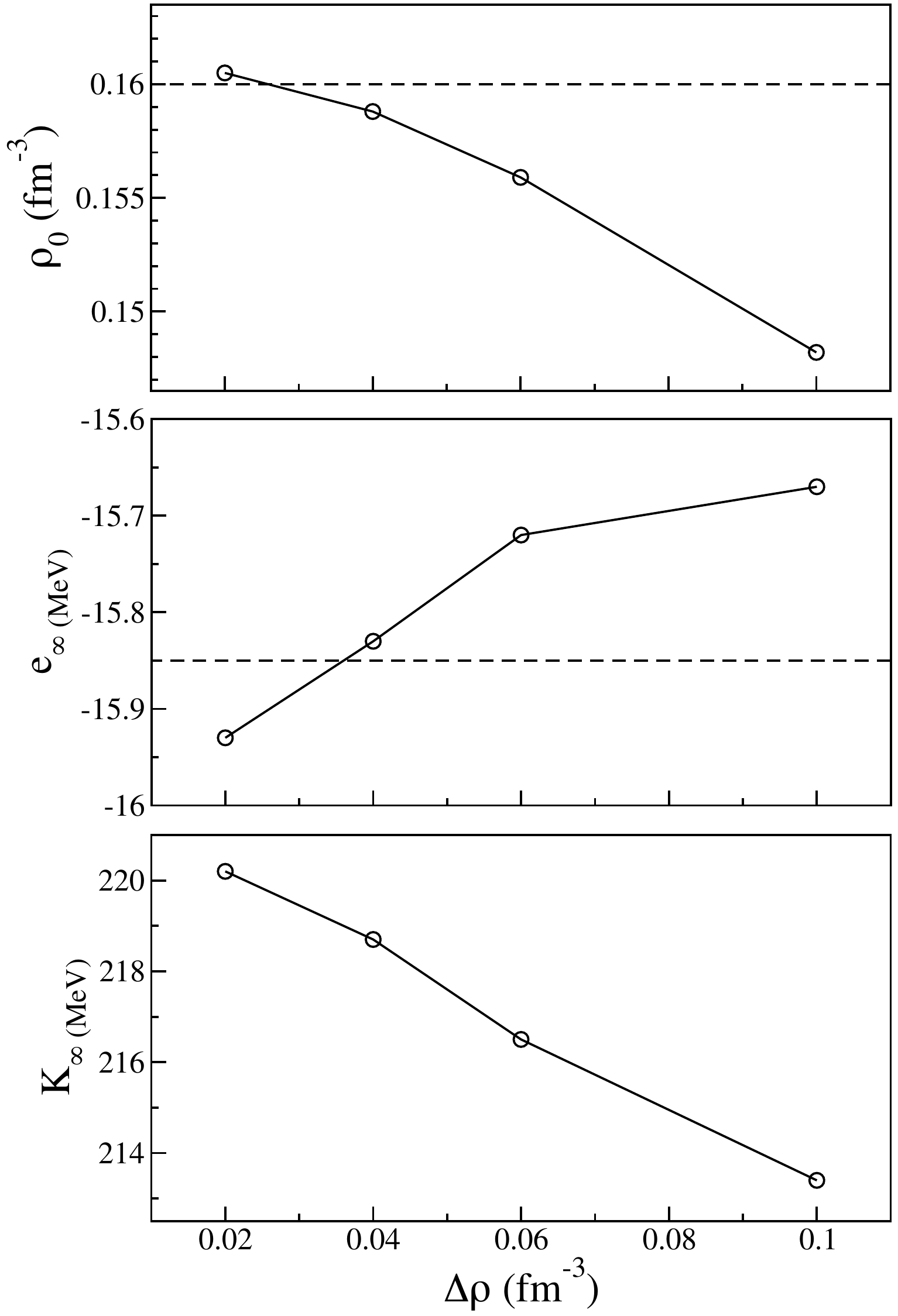}
   \caption{\label{fig:range} Saturation density $\rho_0$, energy per nucleon at saturation $e_\infty$, and incompressibility $K_\infty$ of symmetric infinite nuclear matter as a function of the density range of the fit region (see text). The dashed lines indicate the values of the pseudodata used in the fitting protocol [see Eqs.~(\ref{eq:pseudodatarho}) and (\ref{eq:pseudodataE})].               } 
\end{figure}
{
While in Ref. \cite{Guichon2006}, the  fit region for the density was chosen as $[0-0.2]$ (fm$^{-3}$), Wang {\it et al} used a Taylor expansion around saturation to determine the Skyrme parameters from QMC.
Here, various sizes of the fit region have been considered. 
Figure \ref{fig:range} shows that varying the fit region for the density shifts the energy per nucleon and the density at saturation. Although both $\Delta \rho =0.02$ and 0.04~fm$^{-3}$  provide reasonable values for the properties of the symmetric infinite nuclear matter, $\Delta\rho=0.04$~fm$^{-3}$, corresponding to a fit region of $0.12-0.2$~fm$^{-3}$, provides a better prediction of the $^{208}$Pb binding energy (see Table \ref{tab:fit_sensitivity}). The latter is then chosen to determine SQMC parameters. }

\bibliography{References}
\bibliographystyle{unsrt}
\end{document}